\documentclass[12pt,preprint]{aastex}
\slugcomment{submitted to ApJ}

\shorttitle{White Dwarfs cooling}
\shortauthors{Prada Moroni \& Straniero}

\begin{document}
\title{Calibration of White Dwarf cooling sequences: theoretical uncertainty}

\author{Pier Giorgio Prada Moroni}
\affil{Dipartimento di Fisica Universit\`a di Pisa, 56127 Pisa, Italy\\ INFN, Sezione di Pisa, 56010 Pisa, Italy\\
Osservatorio Astronomico di Collurania, 64100 Teramo, Italy }
\and
\author{Oscar Straniero}
\affil{Osservatorio Astronomico di Collurania, 64100 Teramo, Italy}

\begin{abstract}
White Dwarf luminosities are powerful age indicators, whose calibration should 
be based on reliable models.
We discuss the uncertainty of some chemical and physical parameters 
and their influence on the age estimated by means of white dwarf cooling sequences.
Models at the beginning of the white dwarf sequence
have been obtained on the base of progenitor evolutionary tracks 
computed starting from the
zero age horizontal branch and for a typical halo chemical composition (Z=0.0001, Y=0.23).
The uncertainties
due to nuclear reaction rates, convection, mass loss and initial chemical composition
are discussed. Then, various cooling sequences for a typical white dwarf mass (M=0.6 M$_\odot$) 
have been calculated under different assumptions on some
input physics, namely: conductive opacity, contribution of the ion-electron interaction
to the free energy and microscopic diffusion. Finally we present the evolution 
of white dwarfs having mass ranging between 0.5 and 0.9 M$_\odot$.
Much effort has been spent to extend the equation of state down to the low temperature
and high density regime. An analysis of the latest improvement in the physics of white
dwarf interiors is presented.    
We conclude that at the faint end of the cooling sequence ($log L/L_{\odot} \sim -5.5$) 
the present overall uncertainty on the age is of the order of 20\%, which correspond to
about 3 Gyr. We suggest that this uncertainty could be substantially reduced by improving 
our knowledge
of the conductive opacity (especially in the partially degenerate regime) and by fixing the 
internal stratification of C and O.
\end{abstract}

\section {Introduction}
In the last few years a growing amount of high-quality
data concerning white dwarfs (WDs), both from space as well as from ground
based telescopes, have
been produced. In particular,
the discovery of WD sequences in Globular Clusters (Paresce, De Marchi, \& Romaniello 1995,
Richer et al. 1997) and the
observation of their faint ends in old Open Clusters (Von Hippel, Gilmore, \& Jones 1995, Von Hippel \& Gilmore 2000)
provide new tools for the study
of the history of both the halo and the disk of the Milky Way.
The possible
use of the observed WD sequences as age indicators for a
variety of Galactic components is particularly promising.
As firstly shown by Mestel (1952), the luminosity of a WD is
largely supplied by its thermal energy content, so that the cooling time scale
is inversely proportional to a certain power of the luminosity. Thus,
comparisons between theoretical estimations of the age-luminosity relation for WDs and
the observed cooling sequences may be used
to derive stellar ages.  On the other side,
an accurate comparison between the predictions of the theoretical models
with specific observations of WDs could provide a better comprehension
of the physical properties of high density matter. In this context
the recent development of the astereoseismology applied to WDs allows to check
the internal structure of these compact objects (Bradley \& Winget 1991, Bradley, Winget, \&
Wood 1992, Clemens 1993, Metcalfe, Winget, \& Charbonneau 2001).

White dwarfs are the final destiny of the evolution of low and intermediate mass stars.
The majority of the presently observed WDs are post-Asymptotic Giant Branch stars (AGB).
The core of an AGB star is made of ashes of the He-burning,
namely a mixture of primary carbon and oxygen, with a minor amount of secondary $^{22}$Ne\footnote
{The original C,N and O nuclei are firstly converted into  $^{14}$N, during the H-burning,
and, later on during the He-burning, into $^{22}$Ne  via the chain
$^{14}$N$(\alpha,\gamma)^{18}$F$(\beta^+,\gamma)^{18}$O$(\alpha,\gamma)^{22}$Ne. Thus,
the abundance (in number) of $^{22}$Ne in the CO core of a WD roughly correspond to the
abundance (in number) of the original CNO.}.
During this evolutionary phase the core mass increases,
due to the accretion of fresh material processed by the He-burning shell.
In the meantime, a huge mass loss erodes the
H-rich envelope until it is reduced down to a critical value (see e.g. Castellani, Limongi, \& Tornamb\'e 1995),
and the star departs from the AGB. About $98-99\%$ of the total mass of the resulting WD is
the CO core builded up during both the central and shell He-burning phases.
M$_{up}$ is the minimum stellar mass for
which a carbon ignition occurs before the onset of the AGB and set
the upper mass limit for the progenitors of CO WDs.
Current stellar evolution theory predicts that
the value of M$_{up}$ varies between 6.5 and 7.5
M$_{\odot}$, for metallicity ranging between
Z=0 and 0.02 (see Cassisi, Castellani, \& Tornamb\'e 1996 and Dominguez et al. 1999 for recent reviews
of these calculations).
According to a well established theoretical scenario, single stars having mass lower than M$_{up}$
end their life as
CO WDs with masses ranging
between about 0.55 and 1.05 M$_{\odot}$ (see e.g. Dominguez et al. 1999).
The theoretical initial-final
mass relation, as derived from stellar evolution models, roughly agrees with the
few available measurements of WD masses in nearby Open Clusters
(Weidemann \& Koester 1983, Weidemann 1987, 2000, Herwig 1997).
However, it should be recalled that  stellar
rotation could induce the formation of more massive CO WD (Dominguez et al. 1996).

Few WDs may have an He-rich core.
In principle, stars having initial mass particularly low (namely M$\le$0.5 M${_\odot}$)
never attain temperatures high enough for the He-burning. After the main sequence,
they develop
a degenerate He core and their final fate should be a low mass He-rich compact remnant.
However, owing to the very long evolutionary time scale of these stars
(much longer than the Hubble time), we can definitely exclude that
the present generation of WDs could include their remnants.
Nevertheless, low mass He-rich WDs may result from the deviated evolution of more massive
progenitors when, under particular circumstances,
a complete envelope removal takes place during the first ascent of the red giant branch.
For example, a stripping of the envelope may be caused by a close encounter with another star
(Livne \& Tuchman 1988)
or by Roche lobe overflow in close binary systems (Kippenhahn, Koll, \& Weigart 1967, Kippenhahn,
Thomas, \& Weigart 1968, Iben \& Tutukov 1986).

It has been also supposed that massive WDs could be generated by stars having an initial
mass slightly larger than M$_{up}$  (Dominguez, Tornamb\'e, \& Isern 1993,
 Ritossa, Garcia-Berro, \& Iben 1996, 1999; Garcia-Berro, Ritossa, \& Iben 1997).
In this case, in fact, an highly degenerate O-Ne-Mg core
is left by the carbon burning. Then, after a rather normal AGB phase,
eventually characterized by an important mass loss,
these stars could terminate their life as massive ONeMg WDs (M${\sim}1.2-1.3$ M$_{\odot}$).

In this paper, we confine our analysis to the CO WDs having an H-rich envelope, which
are the most common type of these compact remnants (the so called DA).
Let us note that the physical evolution of the internal structure of a white dwarf
appears rather simple when
compared to other stellar evolutionary phases.
The residual nuclear burnings provide
only a negligible contribution to the total luminosity. The highly degenerate core is
almost incompressible, so that only the contraction of the most external layers may still
contribute to the release of
gravitational energy. Only the brightest WDs (logL/L$_{\odot}\gtrsim -1.5$)
suffer a non-negligible energy loss through neutrino production.
The cooling time is then determined by the rate of temperature decrease, which depends
on the efficiency of the energy transport from the hot core through
the thin opaque envelope. A very efficient thermal conduction, due to the highly
degenerate electrons, takes place in the whole CO core that rapidly
becomes almost isothermal. On the contrary,
the more external layers of the WD (i.e. an He-rich mantel and, eventually, an H-rich envelope,
where electron are initially only partially or non degenerate)
are efficient insulators and regulate the energy loss rate. Basically,
the time scale of the WD evolution is controlled by the heat capacity of the
CO core and by the opacity of the external wrap. Thus, the main problem of this studies
concerns the accurate evaluation of these two physical ingredients and no
sophisticated algorithms for the computation of the stellar structure is, in principle, required. 
Nonetheless, we believe that the best approach should be that based on a full
evolutionary code calculation, which allows a coherent
description of both the progenitor and the WD evolution and may account for all the
contributions to the energy production and transport. 

In this paper we critically re-analyze the influence of the assumed theoretical scheme
on the predicted cooling time. In such a way,
we try to evaluate the present uncertainty
of the cosmo-chronology predictions based on WD models.
A rough idea of the present level of uncertainty may be obtained by comparing some
recently published theoretical cooling sequences for H-rich atmosphere WD (see figure 1).
One may recognize sizeable differences, in particular near the faint end of the cooling sequence.
They are likely due to different evaluations
of the basic ingredients of the theoretical cooking:
the specific heat in the core, the radiative and the conductive opacities, the
treatment of the external convection and the like; but they could be also partially attributed
to differences in the
initial models. We argue that the origin of these uncertainties should
be identified before using cooling sequences as cosmic clocks.
Owing to the complexity of the
matter, our investigation is far from being exhaustive. Many previous works have addressed
the question of the reliability of the input physics used in WD models computation. In some case
we do not repeat their analysis. For example, we do not address the question of the chemical
segregation occurring in the core when the crystallization of the Oxygen component takes place,
because this argument has been exhaustively discussed in the recent literature
(Segretain et al. 1994, Salaris et al. 1997, Montgomery et al. 1999, Isern et al. 1997, 2000).
In some case we extent previous investigations up to the faint end of the cooling sequence, as
in the case of the effects of microscopic diffusion.
In the following section we illustrate the input physics used to compute the various
cooling sequences. In the third section we discuss the uncertainties related to the
progenitor evolution that determine the internal stratification of the core and the size
of the thin external layers: the He-rich mantel and the H-rich envelope.
Section four illustrates the dependence of the cooling time scale
on some physical ingredients (EOS, opacity and the like). A final discussion follows.

\section{Input physics}
Our WD models have been obtained by means of a full evolutionary code. In particular we
have used the FRANEC in the version described by Chieffi \& Straniero (1989).
The input physics have been
completely revised in order to account for the peculiar conditions developed in WD interiors.

\subsection{Opacity}
We have adopted the radiative opacities of OPAL (Iglesias
\& Rogers 1996)  for high temperature (log$T[K]> 4.0$) and
the Alexander \& Ferguson (1994) molecular opacities for
the low temperatures (log$T[K]\leq 4.0$). The conductive opacities have been derived from
the work made by Itoh and coworkers (Itoh et al. 1983, Mitake, Ichimaru, \& Itoh 1984, Itoh, Hayashi, \& Kohyama 1993).
Additional models have been obtained by using the Hubbard \& Lampe (1969) prescriptions
and the table provided by A. Potekhin (see Potekhin et al. 1999).
As it is well known (see e.g. Mazzitelli 1994, D'Antona \& Mazzitelli 1990), there is a 
region in the $T-\rho$ plane not covered by the OPAL radiative opacities and where the 
electron conductivity is not yet dominant. Thus one has to extrapolate the radiative 
opacities in order to fill this gap. The problem is alleviated by the
 use, for the lower temperature, of the Alexander \& Ferguson (1994)
 opacities, which extend at larger densities with respect to the OPAL.
Present models have been computed adopting 
a linear extrapolation of the radiative opacity beyond the upper density provided by OPAL.
We have tested the reliability of this choice by changing the extrapolation method. Negligible
effects on the cooling sequences have been found.

\subsection{Nuclear reaction rates and neutrinos}
Nuclear reaction are almost extinct during the WD evolution, but they
play a relevant role in determine the chemical structure of the model
at the beginning of the cooling sequence.
We have used the rates tabulated by the NACRE collaboration (Angulo et al.
1999),
except for the $^{12}$C$(\alpha,\gamma)^{16}$O. During
the He-burning this reaction competes with the $3\alpha$ and regulates the final
C/O ratio in the core (Iben 1972). The great influence on the resulting WD evolution has
been deeply investigated (D'Antona \& Mazzitelli 1990; Salaris et al. 1997).
We have used two different rates, namely the one reported by
Caughlan et al. (1985) and that of Caughlan \& Fowler
(1988). Note that the difference between these two compilation may be roughly considered
as representative of the present experimental
uncertainty level (see e.g. Buchmann 1996).
Nuclear reaction rates are corrected for the weak and intermediate electron
screening by using the prescriptions of
Graboske et al. 1973 and  De Witt, Graboske, \& Cooper 1973, and for the strong
screening by Itoh, Totsuji, \& Ichimaru 1977 and Itoh et al. 1979.

The efficiency of neutrino emission processes
are taken from: Haft, Raffelt, \& Weiss 1994 (plasma neutrinos),
Itoh et al. 1989 (photo and pair neutrinos).

\subsection{Equation of state}
As recognized
long ago (Salpeter 1961) Coulomb interactions play
a relevant role in WD interior, finally leading to crystallization of the stellar core
(Lamb \& Van Horn 1975).
Thus, a detailed treatment of the thermodynamic behavior of both liquid and solid
matter is a
necessary physical ingredient to study the evolution of cold WDs.
The high pressure experienced in the whole core ensures the complete ionization of
carbon and oxygen.
Thus, we have updated and extended the EOS for
fully ionized matter described by Straniero (1988).
In particular, we have revised the treatment of the electrostatic corrections
up to the liquid-solid transition and beyond.
The free energy in the fluid phase can be written as
$$
F_L=F_i^{id}+F_e^{id}+F_i^{ex}+F_e^{ex}+F_{ie}
$$
where $F_i^{id}$ and $F_e^{id}$ are respectively the contribution of the
ideal gas of ions
 and electrons. To compute these terms, as in Straniero (1988), we have assumed that ions
follow the Boltzmann distribution, while electrons are
described
by Fermi-Dirac integrals, for an arbitrary degree of degeneracy and
relativistic state.
$F_i^{ex}$ is the excess  of ionic free energy due to ion - ion Coulomb
interactions.
For this contribution we adopted the analytical expression by Potekhin \&
Chabrier (2000)
that in the region of high ionic coupling parameter ($\Gamma= (Ze)^2/(a
k_B T) \geq 1$)
fits the accurate results of recent Monte Carlo simulations  of one
component plasma (De Witt \& Slattery 1999),
while for $\Gamma < 1$ it approaches to the Cohen \& Murphy (1969)
expansion and
reproduces the Debye-Huckel limit for vanishing $\Gamma$  (Landau \& Lifshitz 1969).
The quantum diffraction correction
to ionic free energy has been neglected in the liquid phase.
$F_e^{ex}$ represents the excess electron free energy due to electron -
electron
interactions. For this term we adopted the non relativistic expression by
Tanaka, Mitake, \& Ichimaru 1985.
We neglected the relativistic $F_e^{ex}$. This is a reasonable
approximation
since at high density, when degenerate electrons become relativistic,
the electron
coupling parameter $\Gamma_e \ll 1$.
Finally, for $F_{ie}$, that describes the excess due to ion - electron
interactions,
we adopted the analytical expression given by Potekhin \& Chabrier
(2000).
 As it is well known, the contribution of the ion-ion interactions increases with
density, eventually leading
to crystallization. According to Potekhin \& Chabrier (2000),
the liquid-solid phase transition has been assumed at $\Gamma_c=175$.
In the solid phase ($\Gamma \geq \Gamma_c$) the free energy can be
written as
$$
F_S=F_e^{id}+F_i^{s}+F_e^{ex}+F_{ie}+F_i^{qm}
$$
where the electron terms $F_e^{id}$ and  $F_e^{ex}$ are the same as in
the fluid phase. For the free energy of the ionic crystal $F_i^{s}$
 we used the analytical expression by Farouki \& Hamaguchi (1993)
obtained by fitting numerical models of molecular dynamic.
The contribution of the ion - electron $F_{ie}$ in the solid phase is from
Potekhin \& Chabrier (2000). Finally,
 the term $F_i^{qm}$ represents the quantum correction to the
thermodynamic behavior of the ionic Coulomb crystal.
This contribution is very important in late WD evolutionary phases,
in fact, as the WD cools down, the crystallized CO core reaches a quantum
state (diffraction parameter $\hbar \Omega_P/(k_B T) > 1$, where
$\Omega_P$ denotes the ion plasma frequency). When this
occurs, the ionic contribution to the specific heat decreases as $T^3$, thus
depleting the main thermal reservoir of the star: 
it is called Debye cooling phase. For this term we adopted the analytical
expression described by Stolzmann \& Blocker (2000), which is based on the free energy of a
Coulomb crystal studied by Chabrier, Ashcroft, \& De Witt (1992).
The various thermodynamic quantities for pure carbon and pure oxygen
have been obtained by analytically deriving the corresponding free energy. Finally,
an additive volume interpolation is used to calculate the thermodynamic quantities of
a CO mixture\footnote{We neglect the contribution of $^{22}$Ne.}.

To describe
the outer layers of partially ionized helium and hydrogen, we adopted the
EOS of Saumon, Chabrier, \& Van Horn (1995, SCVH). This EOS requires both high pressure
and low
pressure extensions.
At high pressure (P$>10^{19}$ dyn/cm$^2$), which is in any case large enough to guarantee a
full ionization of H and He, we extended the SCVH EOS
following the same procedure described above for the CO core.
The match between the two EOS is generally good.
At low pressure, we have used a perfect gas (including H, H$^+$, H$_2$, H$^-$, He, He$^+$ and He$^{++}$)
plus radiation. In such a case the classical Saha
equation has been used to derive  the population of the various species.
For each temperature, the precise value of the maximum pressure of the perfect gas
(or the minimum of the SCVH) has been varied in order to guarantee a smooth transition
between the two EOS.

Tables of this EOS are available on the web.

\subsection{Model Atmospheres}
In order to fix the external boundary condition of a stellar model, appropriate
model atmosphere are needed.
In the last decade WD atmosphere theory has significantly improved
(Bergeron, Waesemael, \& Fontaine 1991; Bergeron, Saumon, \& Waesemel 1995;
Bergeron, Waesemel, \& Beauchamp 1995; Saumon \& Jacobson 1999; Bergeron
2000).
As it is well known, in the high density/low temperature atmospheres of an old WD
the emerging electromagnetic flux
significantly departs from the black body spectrum. This is mainly due to the
molecular
 hydrogen recombination and the consequent collision - induced absorption
(CIA) by
 H$_2$ - H$_2$ collisions. CIA is the main source of opacity in the
infrared
and the main cause of increasingly blue color indices for
decreasing effective temperature (T$_e$) in cold WDs (Hansen 1998, 1999; Saumon \& Jacobson 1999;
Bergeron 2000).
Detailed model atmospheres are also crucial ingredient to transform
the theoretical
quantities (L, T$_e$) into the observational ones (magnitude, color).
In the present computations we have used a solar scaled T-$\tau$ relation 
(see Chieffi \& Straniero for details) 
up to the onset of the cooling sequence. Then, namely when log L/L$_\odot \sim 0$, we
adopt the model atmospheres of
Bergeron, Saumon, \& Wesemael (1995), plus a low temperature extension
including the effects of the CIA, which has been kindly provided us by
P. Bergeron. Since no metals are included in these models, they are consistent with the 
usual hypothesis of complete sorting of the external layers (see next subsection).

\subsection{Treatment of convection and other mixing phenomena}
Convective boundaries are fixed by using the Schwarzschild criterion.
Concerning the progenitor evolution, the extension (in mass) of the convective core during
the central He burning have a great influence on the amount of C (and O) in the most
internal layers of a WD (Imbriani et al. 2001 and Dominguez et al. 2001).
During this phase, we
use the algorithm described by Castellani et al. (1985) to take into account the growth
of the convective core induced by the conversion of He (low opacity) into
C and O (large opacity) and the resulting semiconvection. Breathing pulses, which occur
when the central He mass fraction decreases below about 0.1, are usually neglected
(but see the discussion in section 3.3). Note that any
mechanism that
increases the size of the well mixed region during the final part of the He-burning
(mechanical overshoot, semiconvection, breathing
pulses or rotational induced mixing) leads to a reduction of the resulting amount of C in the
central region of the WD (see e.g. Imbriani et al. 2001).

Theoretical studies (Fontaine \& Michaud 1979, Iben \& MacDonald 1985, 1986, Althaus et al. 2002),
supported by observational evidences (see e.g. Bergeron, Ruiz, \& Legget 1997),
indicates that, as a consequence of the gravitational settling of
heavy elements, DA WDs should have a practically pure H envelope and a pure He mantel.
In addition, as suggested by Salaris et al. (1997),
the carbon and oxygen profiles left by the
He-burning is smoothed by Rayleigh-Taylor instabilities. 
In order to account for the element diffusion, when log L/L$_\odot \sim 0$, 
we adjust the composition of the envelope and the mantel:  
all the residual H is putted on the top of a pure He layer. The total mass of H and He is 
conserved. We also modify the internal C and O profile according the the prescription 
of Salaris et al. (1997).  

Owing to the larger $\Gamma$, oxygen crystallization occurs
when carbon is still liquid. For a certain time, this occurrence
produces an unstable stratification of the liquid phase and, in turn,
 an efficient mixing (Isern et al. 1997, Salaris et al. 1997). This phenomenon is presently
not included in our calculations.

\section{Pre-WD evolution}
First of all one should know the physical and the chemical structure of the star
at the beginning of the cooling sequence. Since  our investigation regards DA white dwarfs,
we need to know the internal stratification of the CO core and the extension (in mass)
of both the He and
H-rich layers.
The formation of the CO core begins during the central He-burning phase and proceeds in
the following AGB phase. In principle, a lengthy computation through a series of recursive
thermal pulses is required, until the
mass loss erodes the envelope and the star becomes a WD (see e.g. Iben \& Renzini 1983).

During the thermally pulsing AGB phase, the mass of the CO core is accreted with
the ashes of the He-burning shell,
while the mass between the two shells is progressively reduced (see e.g. Straniero et al. 1997).
 Thus, it is very important to determine the
duration of the AGB, which is evidently affected by mass loss: the
longer the AGB phase, the larger the
final mass of the CO core and the smaller the mass of the He-rich mantel.
AGB mass loss rate ranges
between 10$^{-8}$ and 10$^{-4}$ M$_\odot$/yr, but it is still unclear how 
it is related  to the changes of the chemical and physical structure occurring in
the cool atmospheres of these giants (see Habing 1996).
In addition, if the WD progenitor is a low mass star (M$<$1.5 M$_\odot$),
the mass loss during the
first ascent to the tip of the red giant branch (RGB) must be also carefully taken into account.
The spread in color of the horizontal branch (HB) stars in
Globular Clusters is commonly interpreted as an evidence of the different mass loss
experienced by similar stars
during the RGB phase: bluer HB stars should have a smaller envelope
than red HB stars. Since the AGB phase of blue (reduced envelope) HB stars
will be significantly shorter than
that of a red HB star, the resulting WD will be different.
In other words, stars with similar initial mass and chemical
composition might produce
white dwarfs of different masses and/or different internal stratification, depending on the
mass loss history.

Not only the mass, but also the initial composition of a star determines its evolution and,
possibly, its final fate. A detailed descriptions of the internal C and O profiles in WDs
generated by stars with mass ranging between 1 and 7 M$_\odot$ and metallicity 0$\le$Z$\le$0.02
may be found in Dominguez et al. (2001).

This variety of possible progenitor evolutions could frighten the reader.
Our scope is to quantify the uncertainty on the cooling time of WD, for different
galactic components. Basing on some numerical experiments, we have evaluated
the influence of the pre-WD history on the age calibration.
Starting from zero age He-burning stellar structures
(i.e. an He core surrounded by an H-rich envelope, in which
 a suitable mixture of Helium and metals is adopted),
we have computed the evolutionary sequences up to the WD stage.
At the beginning of the cooling phase ($logL/L_{\odot}\sim0$) the envelope has been completely sorted in a
  pure He buffer and a pure H outer envelope mimicking the
  effect of the very efficient gravitational settling.
Table 1 reports the initial parameters of the models and the properties of the resulting
white dwarfs,
namely: identification label of the model,
initial mass (M$_{ZAHB}$), metallicity (Z),  initial He-core mass (M$_{He}$),
final mass (M$_{WD}$), final amount of He ($\Delta$M$_{He}$),
final amount of H ($\Delta$M$_{H}$), central carbon mass fraction (X$_C$),
mass of the homogeneous central region (Q=M$_{flat}$/M$_{WD}$, see below for a definition
of this parameter), WD age when log L/L$_\odot$=-5.5.

\subsection{Comparison between different evolutionary tracks leading to the same WD mass}
Let us firstly discuss  models B, B1, B2.
The total mass has been maintained constant (M=0.6 M$_{\odot}$) during the whole evolution
for both case B and B1. They differ for the initial mass
of the H-rich envelope, namely: $10^{-3}$ and $2 \cdot 10^{-2} M_{\odot}$ respectively for B and B1.
The third case (B2) has been computed starting from a larger mass (M=0.64 M$_{\odot}$) and
imposing a constant mass loss ($3 \cdot 10^{-8}$ M$_{\odot}/yr$)
since the onset of the thermally pulsing phase. In this case, the initial
He-core mass is in agreement with recent evolutionary computation of H-burning models of
low mass stars (see e.g. Straniero, Chieffi, \& Limongi 1997).
Note that the initial total mass has been chosen to obtain
a final (WD) mass similar to the (constant) one of case B and B1.
A low metallicity (Z=$10^{-4}$), adequate for old halo WD progenitors, has been
chosen for all these three models.
Figure 2 illustrates the evolutionary tracks in the HR diagram.
The central He-burning
phase is similar in the three cases. At the central He exhaustion, model B1 evolves toward the red
part of the HR diagram and develops the typical double shell structure of an AGB star. When the
faster He-burning shell get closer to the H-burning one, it dies down and the
 star enters in the thermally
pulsing phase (for a more detailed description see e.g. Iben \& Renzini 1983).
After two major thermal pulses, the envelope mass is reduced down to about 0.001 M$_\odot$; then,
the star leaves the AGB. During the post-AGB, a further (last) thermal pulse takes place;
the star re-expand and comes back to the
asymptotic giant branch. Finally, after a second horizontal crossing
of the HR diagram, both shells definitely die down and
the model settles on the WD cooling sequence. 

Owing to the very small envelope mass, after the central He exhaustion, model B
remains in the blue side of the
HR diagram, so skipping the classical AGB phase. When the outgoing
He shell approaches the H-rich layer it dies down. Note that the H-burning shell
has been practically inactive
up to this moment. The following contraction of the star induces
some irregular and short H-ignitions.
Finally this model attains its cooling sequence.

In model B2 the duration of the AGB and the final WD mass are mainly determined by the
 assumed mass loss rate, which is, in any case, quite typical for faint TP-AGB stars.
This models experience 8
thermal pulses; then, the envelope is reduced down to a certain critical value
and the
model moves toward
the blue side of the HR diagram (see figure 2). Note that the mass loss
has been stopped when the effective temperature becomes larger than 7000 K.
At variance with case B1, no more thermal pulses are found
in the post-AGB phase.

The resulting profiles of carbon and oxygen in the core of the three final WD models are shown
in figure 3.
The most internal region is produced during the central He-burning and its flatness is
a consequence of the
convective mixing occurring in the stellar core. As noted by Salaris et al. (1997),
the region immediately outside the flat profile left by the central convection
presents a positive gradient of mean molecular weight. The consequent Rayleigh-Taylor
instability induces a mixing and a further little reduction
of the central carbon. In column 8 and 9 of table 1 we have reported the central C mass fraction
and the mass of the homogeneous central region
(in unit of the total WD mass) as they should result after the action of
this Rayleigh-Taylor instability (Q=M$_{flat}$/M$_{WD}$).
The growing C profile in the more external layer
is produced during the AGB,
when the He-burning shell advances in mass toward the stellar surface.
The carbon peak at the external border of the CO core reflects the incomplete He-burning
left by the last thermal pulse. The three chemical structures
are rather similar. In particular the amount of carbon (oxygen) in the center of the WD and the
extension of the inner flat region are practically the same.
The most relevant differences is the size (in mass)
of the
He-rich mantel (see column 6 in table 1).  In model B1 the thickness of the mantel is
about half of that found in case B. This is the memory of the last thermal
pulse experienced by model B1 in the post-AGB phase, when the H-burning shell was almost
extinct.

The amount of hydrogen left in the envelope of the WD is practically
the same in the three models.
This confirms the fact, already demonstrated in many papers, that
for each metallicity and core mass, it exist a maximum
mass of the H-rich envelope for a WD (Iben 1982, Fujimoto 1982a,b,
Castellani, Degl'Innocenti, \& Romaniello 1994, Piersanti et al. 2000).
When the H mass exceeds such a limit,
the H-burning is efficiently activated at the base of the envelope. It
drives the expansion of the envelope leading to the formation of a
giant star rather than to a white dwarf.

The cooling sequence obtained starting from the three pre-WD models illustrated
in this subsection are shown in figure 4. They are very similar, with a little difference
(less than 0.5 Gyr)
at the faint end. In particular, the cooling of model B1 is a bit
faster because the smaller He mantel.
Note that these three models reach the WD stage through a very different path
and they may be considered as representative of the possible
changes occurring during the progenitor evolution.
In summary, the knowledge of the cooling time scale at the faint end of the WD sequence (namely
at log L/L$_\odot$=-5.5) is limited
by a 3-4\% error as a consequence of the uncertainty on the leading parameters characterizing 
the pre-WD history.

\subsection{Influence of the initial metallicity}
In many recent studies of cosmochronology (Salaris et al. 1997, Richer et al. 2000)
WD models based on progenitors having solar chemical composition have been used.
This is certainly adequate to study the age of the disk and its components (Open Clusters).
 However, when the main goal are the oldest component of our Galaxy, halo or
Globular Cluster stars, low metallicity models are more appropriate.
In this subsection we would try to evaluate the influence on the cooling sequence
of the progenitor chemical composition.
For this reason we have computed an additional model producing a WD having M=0.6 M$_\odot$, but
with a solar metallicity, namely Z=0.02.

The properties of this model are reported in the fourth row of table 1 (model B3). A comparison
with model B (Z=0.0001) reveals only small differences. As already discussed by Dominguez et al. (2001),
 the central carbon abundance is slightly lower
at large metallicity; they found a maximum 10\% variation when the metallicity is varied
between 0.0001 and 0.02. Note that
the extension of the innermost homogeneous region decreases by increasing the metallicity
(see the values reported in column 9 of table 1). As expected, since the
H-burning is more efficient at large metallicity (because the larger amount of CNO), the
mass of the H-rich envelope is significantly reduced when the star attains the cooling sequence.
In spite of this difference, at the faint end model B3 is only 2\% younger than model B.
The cooling sequences obtained for the two metallicities are compared in figure 5.

\subsection{Internal Stratification: $^{12}$C$(\alpha,\gamma)^{16}$O
 and turbulent convection}
The amounts of carbon and oxygen left in the core of a WD have a great influence on its
cooling rate
(D'Antona \& Mazzitelli 1990, Salaris et al. 1997) and plays an important role in determining the
observable outcome of a type Ia supernova (Dominguez et al. 2001).
The chemistry of a WD is determined by the competition of the two major
nuclear reactions powering the He burning, namely  the 3$\alpha$ (the carbon producer) and
$^{12}$C$(\alpha,\gamma)^{16}$O (the carbon destroyer).
As deeply discussed by Imbriani et al. (2001), the final C/O not only depends on the rate of
these two reactions, but it is significantly influenced by the efficiency of the convective
mixing operating near the center of an He burning star.
The pre-WD models discussed so far, has been obtained by
neglecting any source of extra-mixing, like mechanical 
overshoot, possibly occurring at the external edge of the convective core.
We have also suppressed the so called breathing pulses, which take place 
when the central mass fraction of He is reduced down to $\sim$0.1
(see Castellani et al. 1985 and references therein).
As shown by Imbriani et al. (2001), the amount of carbon left in the center is almost insensitive
to the occurrence of the convective overshoot, but it is strongly influenced by
any additional mixing occurring in the final
part of the He-burning, when the carbon previously accumulated by the 3$\alpha$
is efficiently converted into oxygen via the
$^{12}$C$(\alpha,\gamma)^{16}$O reaction. This is exactly the job done by
breathing pulses (BPs).
In this subsection we study the combined effects of a change
of the $^{12}$C$(\alpha,\gamma)^{16}$O rate and the inclusion of BPs.  
Models B to B3 have been obtained by using a high rate for the  $^{12}$C$(\alpha,\gamma)^{16}$O
reaction (namely, the one reported by Caughlan et al. 1985, hereinafter C85).
As it is well known, the adoption of a low rate leads
to a significantly larger central abundance of carbon.
Model B-low has been
obtained by adopting the rate reported by
Caughlan \&  Fowler (1988, hereinafter CF88), which is about 0.4 times the rate of C85 at the
relevant temperatures (around $2 \times 10^8$ K). Recent experimental determinations of this
rate (Buchmann 1996 and 1997, Kunz et al. 2001, 2002) seem to confirm that the actual value should be
intermediate between CF88 and C85.
Then, we have also repeated these two calculations by including the effect of the
breathing pulses (model B-BP and B-low-BP). The internal chemical
profiles of
models B-low  and B-low-BP are compared in figure 6, while the influence on the
cooling sequence is illustrated in figure 7.
In synthesis, the whole uncertainty (convection plus nuclear reaction rate)
due to the He-burning phase induces a
9\% uncertainty in the estimated cooling time at the faint end of the WD sequence
(see the values reported in the last column of table 1).
It is worth noting that this is only a lower limit of the uncertainty caused by the
possible variation of the internal composition. In fact, as discussed by 
Isern et al. (1997) and  Montgomery et al. (1999), the delay of the cooling due to the
chemical segregation upon crystallization is larger when $X_C \sim X_O$. Therefore,
the occurrence of this phenomenon should hampers the differences 
between case B-low and case B-BP. A more quantitative estimation may be deduced from
the results reported in table 2 of Montgomery et al.. Note that our B-low model have an internal 
stratification that roughly correspond to their 50:50 case, while our model B have a composition profile
similar to the one of Salaris et al. (1997). Then, according to Montgomery et al., 
element segregation should increase the age difference between B-low and B-BP of at least 0.6 Gyr, thus
bringing the whole uncertainty on the age at the faint end of the cooling sequence
up to 13-14\%.

\subsection{The thickness of the H-rich envelope}
We have already discussed the existence of a maximum value for the mass of
the H-rich envelope of a WD. However, the actual amount of H at the top of a WD
could be substantially lower.  In fact, stellar wind may continue
during the post-AGB phase so that the H-rich envelope could be strongly reduced and even
completely
lost. It is well known that in an important
fraction (about 10-20 $\%$) of the whole WD population
the H-rich envelope is practically absent (DB-WDs). The minimum detectable amount
of H is very low (about $10^{-10}$ M$_\odot$). The post-AGB mass loss
is the easiest explanation
of the lack of H in the atmosphere of the DB WD. An alternative possibility was
advanced by Iben \& Renzini (1983). They suggested
that the progenitors of
the  DB could have suffered a last thermal pulse during the post-AGB, when the H-burning shell is
almost extinguished. In this condition, the convective shell generated by the thermal pulse could
completely mix the He and the H-rich layers. Then, the H dredged down at high temperature is
rapidly consumed. We recall that a similar
post-AGB thermal pulse has been encountered in our model B1, but no evidence of a deep H mixing
has been found (in this particular case).

D'Antona \& Mazzitelli (1989) and Tassoul et al. (1990) have investigated the WD properties
by changing the amount of H in the envelope. However, due to the lack of reliable opacity and EOS,
they didn't extent the computation down to the faint end
of the cooling sequence. Then, we have computed two additional cooling sequences by artificially
reducing the H-rich layer of model B, namely: $\Delta$ M$_{H}=3.12 \cdot 10^{-5}$ and
$2.32 \cdot 10^{-6} M_{\odot}$ for model B-LH and B-VLH, respectively (row 5 and 6 of table 1).
The comparison between these two models and
model B is shown in figure 8.
As expected, the cooling is faster in models with a smaller envelope.
The age reduction at log L/L$_\odot=-5.5$ would be as large as 1 Gyr if the residual
mass of the H-rich envelope was of the order of $10^{-10}$ M$_\odot$.

\section{The physics of WD interior}
In this section we address the uncertainty caused by some input physics directly affecting
the WD evolution.

\subsection{Conductive Opacities}
Degenerate electrons dominates the heat conduction in white dwarf interiors. Two different
tabulations of the thermal conductivity by electrons are commonly used in the
computation of WD models, namely: the one presented  by
Hubbard \& Lampe (1969, hereinafter HL69) and those by Itoh and coworkers (Itoh et al. 1983, Mitake et al.
1984, Itoh et al. 1993, hereinafter I93).
HL69 considered partially degenerate electrons scattered by ions. They also included
the effect of the electron-electron interaction, which is negligible in extreme
degeneracy regime (due to the Pauli exclusion principle), but becomes important if
degeneracy is not strong, as in the external region of a WD.
The main limitation of
this calculation is due to the assumption of
non-relativistic electrons. In practice, the Hubbard and Lampe conductive opacities
are only adequate for
densities lower than $10^6$ g/cm$^3$. Actually, a stronger
limitation of this calculation is often ignored. In fact, in order to smooth out
the discontinuity in the conductive opacity
occurring at the liquid/solid phase transition, the conductive opacities were
artificially and progressively increased for $\Gamma \ge$10.
Let us recall that at the end of the Sixties it was believed that
the phase transition from liquid to solid would occur for rather low values of
the Coulomb coupling parameter ($50 < \Gamma < 100$). This explains the choice of a
particularly low
$\Gamma$ to manage the discontinuity occurring at the phase transition.
Nevertheless, the most recent
studies find that the crystallization begins at $\Gamma \sim 170-180$ (we adopt 175), in
any case well above $\Gamma=10$.
The work made by Itoh and coworkers is a substantial improvement of
the calculation of electron conduction. They treat inter-ionic correlations
and electron-ion interactions by means of an improved theory, based on one-component Monte Carlo
calculation.
However, since they assume fully degenerate
electrons, their opacities are only accurate for large value of the degeneracy parameter.
At lower density (or at larger temperature), I93 probably underestimate the effects of the
electron-electron interactions.

In figure 9 we illustrate the situation in the case of the core of our reference model (B).
The empty area in the left panel shows the allowed region in the case of HL69, while
the right panel refers to I93. The two (almost vertical) series of arrows shows the 
evolution of the physical conditions at the center and at the external boundary of the core.
This figure demonstrate that    
HL69 is not adequate to describe the heat conduction in the major part of the WD
structure, even if
the density is somewhat lower than the relativistic limit.
On the contrary the condition of extreme degeneracy is well satisfied
and the I93 assumption seems appropriated for
the whole CO core. Concerning the more external layers, figure 10 illustrate the case
for the He-rich mantel. As in figure 9, the left and the right panel show the region of validity
for HL69 and I93, respectively. The two series of arrows marks the boundaries of the mantel. 
HL69 should be preferred in the first part of the cooling sequence,
 since electrons may be partially degenerate
while $\Gamma$ is generally rather small.
The situations change toward the faint end of the cooling sequence, during which the
I93 should be more appropriate. 
Note that there exist a thin region in the T-$\rho$ plane in which both calculation are, in principle,
valid. It happens, for example in the case of an He-rich mixture, for $\theta \sim 0.1$. The thick
line in figure 10 marks this condition. However, a comparison along the $\theta$=0.1 line
shows that the HL69 opacities
are generally larger (20-40 \%) than the I93 ones. Such a 
mismatch reveals a more profound difference between the two calculations, which cannot be  
easily understood. 

In figure 11 we report the cooling sequences obtained by using
the HL69 and I93 (case B-HL69 and case B, respectively).
In the former case, we have generated the conductive opacities
by means of the original code provided by Hubbard.
As expected, the lower cooling rate obtained with Hubbard \& Lampe reflects 
the larger value of the conductive opacities. 
Note that the major discrepancies between the two cooling sequences occurs 
at about log(L/L$_{\odot})$=-4, exactly when electron conduction becomes the dominant 
mechanism of energy transportation at the base of the convective envelope. 
As it is well known (see e.g. Tassoul et al. 1990), this occurrence produces
 a temporary reduction of the cooling rate.
Since many authors use a combination of the HL69 and I93, in figure 11
we have also reported an additional model obtained by using I93 
for $\theta<0.1$ and HL69 elsewhere (case B-comb). This test shows that
the discrepancy mainly concerns the conductive opacity in the weakly-degenerate regime.
In conclusion, no ones of the prescriptions widely adopted in the computation of
models of WDs appear adequate to describe the
whole structure of these cool stars. Unfortunately, the substantial disagreement, at
temperature and density
for which both computations should be correct, suggests extreme caution in using combined tables.
Since the differences of these two theoretical prescriptions
imply a 17\% variation in the estimated WD age (at the faint end of the cooling sequence,
see the last column in table 1),
we conclude that the calculation of the conductive opacities (especially in partially
degenerate regime) deserves much attention in view of a reliable
calibration of the WD cooling time.
In this context a recent paper by Potekhin et al. (1999, see also Potekhin 1999)
address this problem, extending the
computation of electron conductivity to the weakly-degenerate regime. Other improvements,
as, in particular, refined ion structure factors and multi-phonon scattering, have been included.
The conductive opacity obtained by Potekhin and coworkers are intermediate between those of HL69
and I93. Nevertheless, as shown in figure 11, the resulting cooling sequence does not substantially differs from the
one obtained by using the I93. On the contrary, the cooling time is significantly smaller then 
those obtained in the model B-HL69 and B-comb. Some properties of the model obtained by means of
the electron conductivity calculated by Potekhin et al. (1999) are reported in table 1 (model
B-Pot).

\subsection{EOS: ion-electron interaction}
Chabrier et al. 2000 compare WD cooling sequences obtained by changing the prescriptions for the 
contribution of the ion-electron interactions to the free energy. In particular they found 
a significant decrease of the cooling time (about 10\%) when the prescription of
Potekhin \& Chabrier (2000) are used in the solid phase instead of the result given by  Yakovlev \& Shalybkov (1989).
As pointed out by Potekhin and Chabrier, the ion-electron 
interactions are generally negligible, except in the solid state at very high density. 
For carbon the ion-electron contribution to the heat capacity 
becomes very important at T=$10^5$ K and $\rho > 10^6$ g cm$^{-3}$ 
but at T=$10^6$ K, $\rho$ must be larger then $10^8$ g cm$^{-3}$, a density never attained in CO
white dwarfs (see figure 8 of Potekhin \& Chabrier). We 
have investigated the overall effect of this contribution to the heat
capacity by computing an additional model (model B-noie) by fully neglecting the ion-electron
interactions. We recall that our "standard" computation are based on the Potekhin \& Chabrier
prescription. The result is shown in figure 12 (see also table 1). This test clearly demonstrates
that the contribution of the ion-electron interaction derived by Potekhin \& Chabrier
produces a negligible effect on the cooling of CO white dwarfs.

\subsection{External Stratification: microscopic diffusion}
Many studies have shown that the time-scale of diffusion is
short enough to sort the external layers of a WD, thus explaining the observed mono-elemental
feature of WD spectra (Fontaine \& Michaud 1979, Muchmore 1984,
Paquette et al. 1986). Many authors have discussed the effects of diffusion
on the WD structure and evolution (see in particular Iben \& MacDonald 1985 and 1986,
Althaus et al. 2002).
All the models presented here have been obtained by assuming a complete sorting of the H and He
layers at the beginning of the cooling sequence ($logL/L_{\odot} \sim 0$).
 Since we are mainly interested in quantify the overall variation of the cooling time,
we have computed a new model in which the chemical stratification of both the envelope and the
mantel
is the one modeled by the nuclear burning and, eventually, by convection (model B-nodif).
We note that the adopted model atmospheres does not include the effect of metals. 
Since the assumed metallicity is quite low (Z=0.0001), the consequent inconsistency should not 
dramatically affect our result.
The comparison between the properties of
this model and those of the {\it standard} one (model B) is illustrated in figure 13
and in table 1.
In the standard case, the larger radiative opacity of the pure H
induces the formation of a deeper convective envelope. As a consequence,
the slow cooling phase, which takes place when
the internal border of the convective layer reaches the region dominated by electron conduction,
is anticipated. In conclusion, at the faint end of the cooling sequence,
the model in which the effect of microscopic diffusion are
neglected is a 11\% younger than the standard model. Note that this is the younger model we
found.

\section{Cooling time and WD mass}
As it is well known the cooling time depends, among the other things, on the WD mass. Figure 14
shows models in the mass range 0.5 - 0.9 $M_{\odot}$ computed under the same prescriptions
adopted for the case B.
In table 2 we have summarized the properties of these models. Isochrones and luminosity
functions based on these models are available on the web.

\section{Summary and Conclusions}
In the last decade cosmochronology based on WDs has become a promising tool
of scientific inquiry thanks to  both the availability of a large amount of
high quality data for cool and faint WDs and the improvement
of the understanding of the high density plasma physics.
However, as demonstrated by the rather large discrepancies
among the recently published theoretical cooling sequences,
a firm calibration of the age-luminosity relation is still not
available.

We have analyzed some of the main sources of
uncertainty affecting the theoretical cooling time.
Physical and chemical parameters characterizing the white dwarfs and
the progenitors evolutions have been revised.

Concerning progenitors, we found that the larger uncertainty is due to
the combined action of convective mixing a nuclear reactions operating
during the central He-burning phase. Both these processes are largely affected by
theoretical and experimental uncertainties. They determine the amount of C (and O)
left in the core of the WD and, in turn, have a great influence on the predicted
cooling rate. A conservative analysis allow us to conclude that the
overall impact of the uncertainties due to the progenitors evolution
on the estimated WD ages at logL/L$_\odot = -5.5$ is of the order of 2 Gyr. We hope that
this uncertainty will be significantly reduced in the next future
thanks to the renewed effort of nuclear physicists in measuring
the $^{12}$C$(\alpha,\gamma)^{16}$O reaction rate at low energy (see Gialanella et al. 2001 and
Kunz et al. 2001).

Concerning WD physics we emphasize the relevance of a reliable description of the electron
conductivity which is the main mechanism of energy transport in WD interiors. Actually, a
consistent part of the large discrepancies found in the comparisons of published cooling
sequences may be due to the adopted conductive opacity. The old Hubbard and
Lampe (1969) conductive opacity are probably overestimated, even in the weakly degenerate
regime. The latest computations by Itoh and coworkers and Potekhin et al. (1999) imply a
substantial reduction of the age. At log$L/L_\odot = -5.5$, we obtain models $\sim 2.5$ Gyr
younger than that obtained with HL69.

Let us conclude by noting that most of the uncertainties discussed in this paper
mainly affect old WDs, but have negligible influence on the age estimated for young
WDs, whose luminosity is larger than $10^{-4}$ L$_\odot$. This implies that the
presently available theoretical scenario may be safely used to estimate
the age of young stellar systems as, for example, the intermediate age Open Clusters
(t$<$3 Gyr).

\noindent {\bf Acknowledgments:}

We wish to thank P. Bergeron for kindly providing us the models atmosphere 
and S. Degl'Innocenti, Luciano Piersanti and Jordy Isern for the many helpful comments.
We thank also S. Shore for the careful reading of the manuscript and the several 
useful discussions.
We are deeply indebted to V. Castellani for continuous encouragements and for a critical review
of the manuscript. This work has been supported by the 
MIUR Italian grant Cofin2000.

{} 
 

\clearpage

\begin{figure}
\plotone{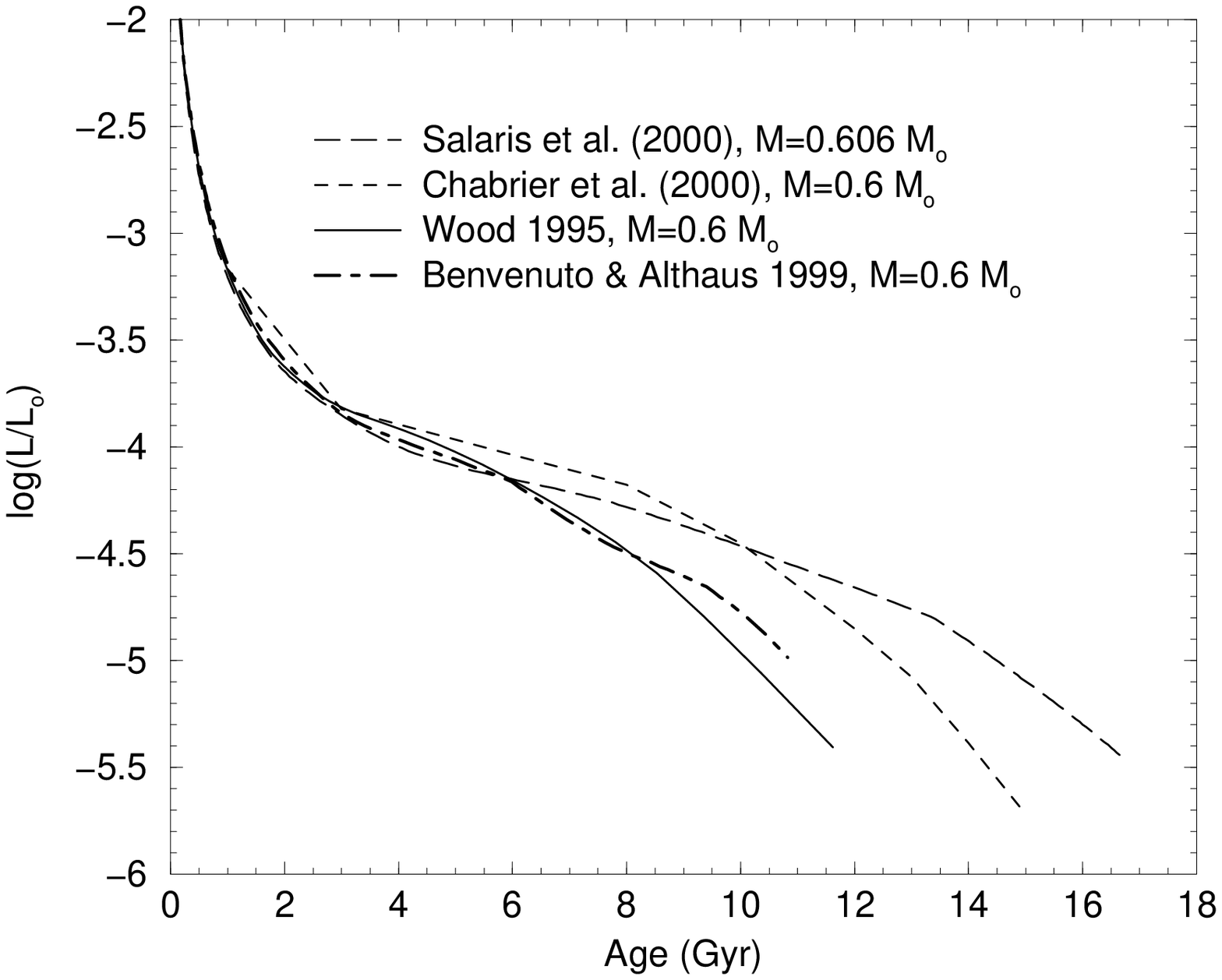}
\caption{Comparison between some of the most recent theoretical cooling sequences
 for a DA WD of M=0.6 $M_{\odot}$: Salaris et al. 2000 
(long dashed line), Chabrier et al. 2000 (dashed line), Wood 1995 (solid line), 
Benvenuto \& Althaus 1999 (dot-dashed linea). All these sequences, except the one
of Chabrier et al 2000,  have been computed 
without including the effect of the chemical segregation occurring upon 
crystallization and by using a full evolutionary code. The Chabrier et al. sequence has been 
derived from static WD models and include chemical segregation. 
\label{fig1}}
\end{figure}

\clearpage

\begin{figure}
\plotone{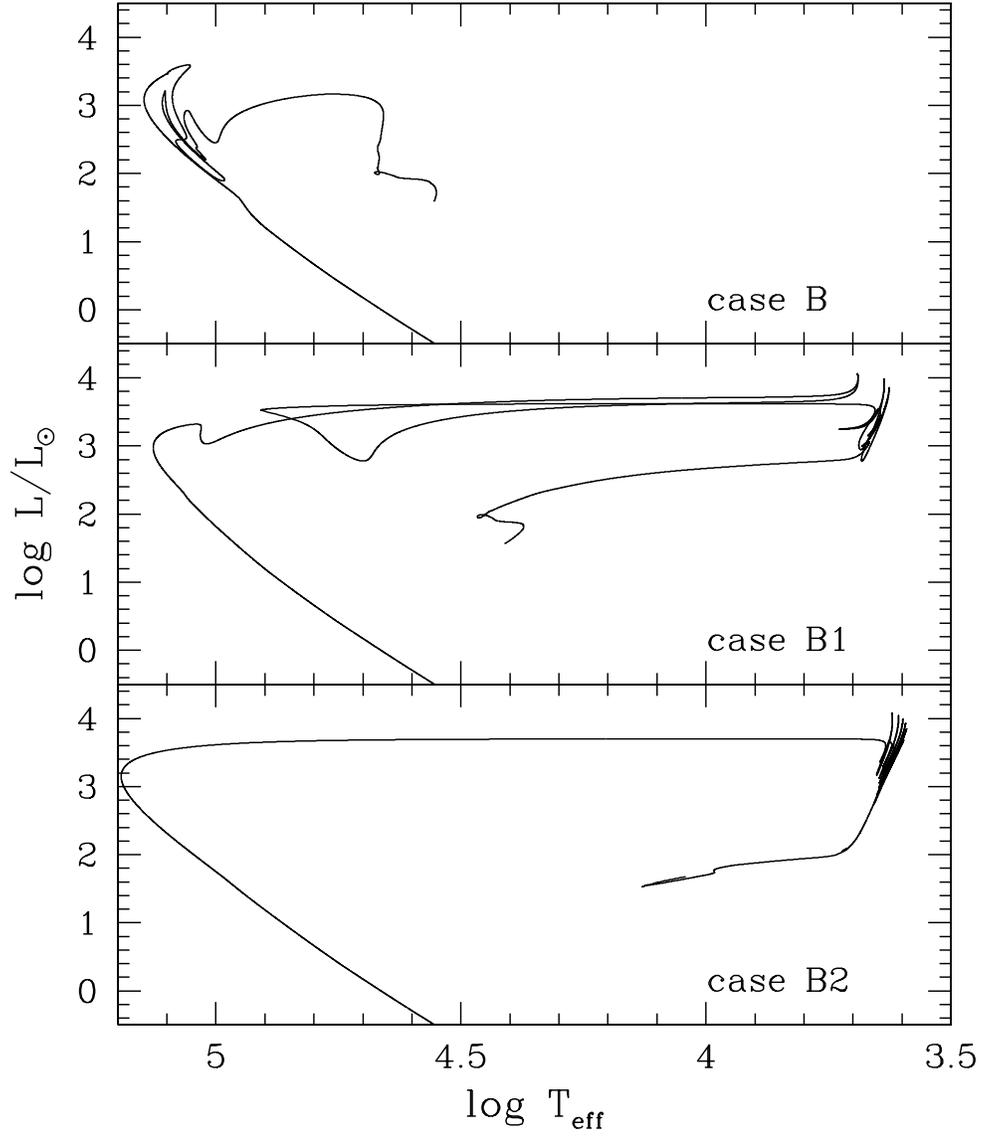}
\caption{Comparison between three different evolutionary tracks leading 
to a 0.6 $M_{\odot}$ WD (models B, B1 and B2 of table 1).\label{fig2}}
\end{figure}

\clearpage

\begin{figure}
\plotone{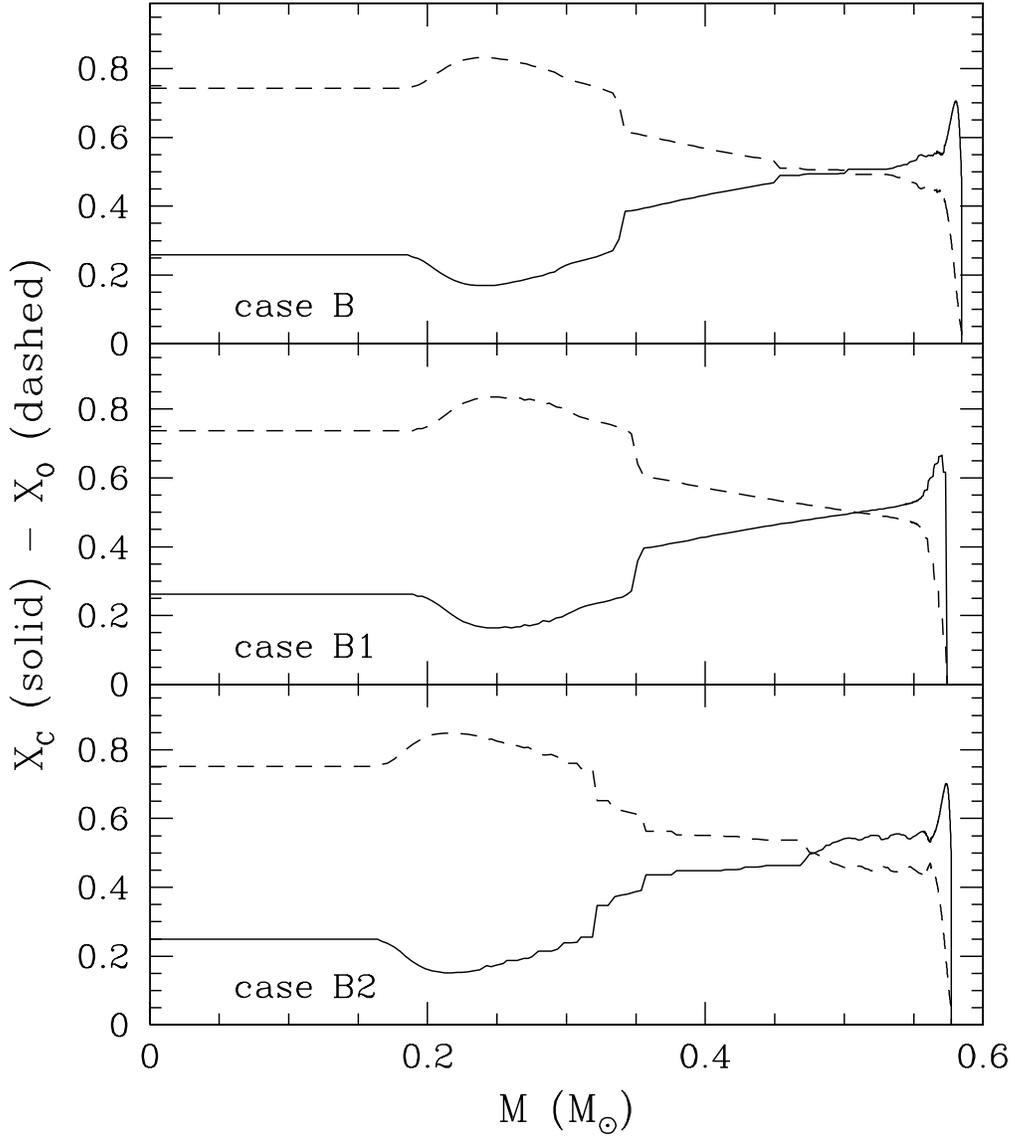}
\caption{Internal profiles of carbon (solid lines) and oxygen (dashed lines)  
for model B (upper panel), B1 (middle panel) and B2 (lower panel), respectively.\label{fig3}}
\end{figure}

\clearpage

\begin{figure}
\plotone{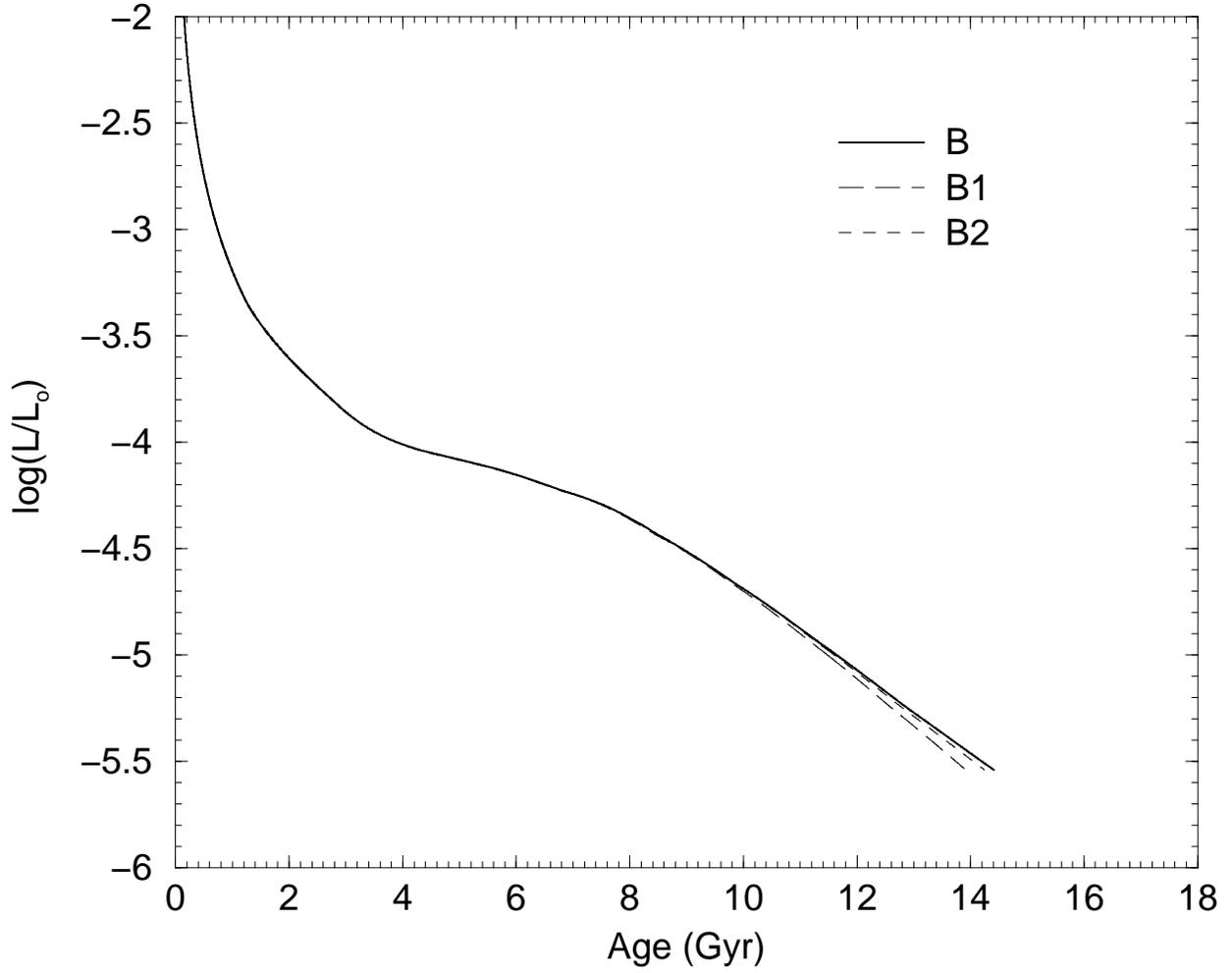}
\caption{Comparison between the cooling sequences for model B (solid line), 
B1 (long-dashed line) and B2 (dashed), respectively.\label{fig4}}
\end{figure}

\clearpage

\begin{figure}
\plotone{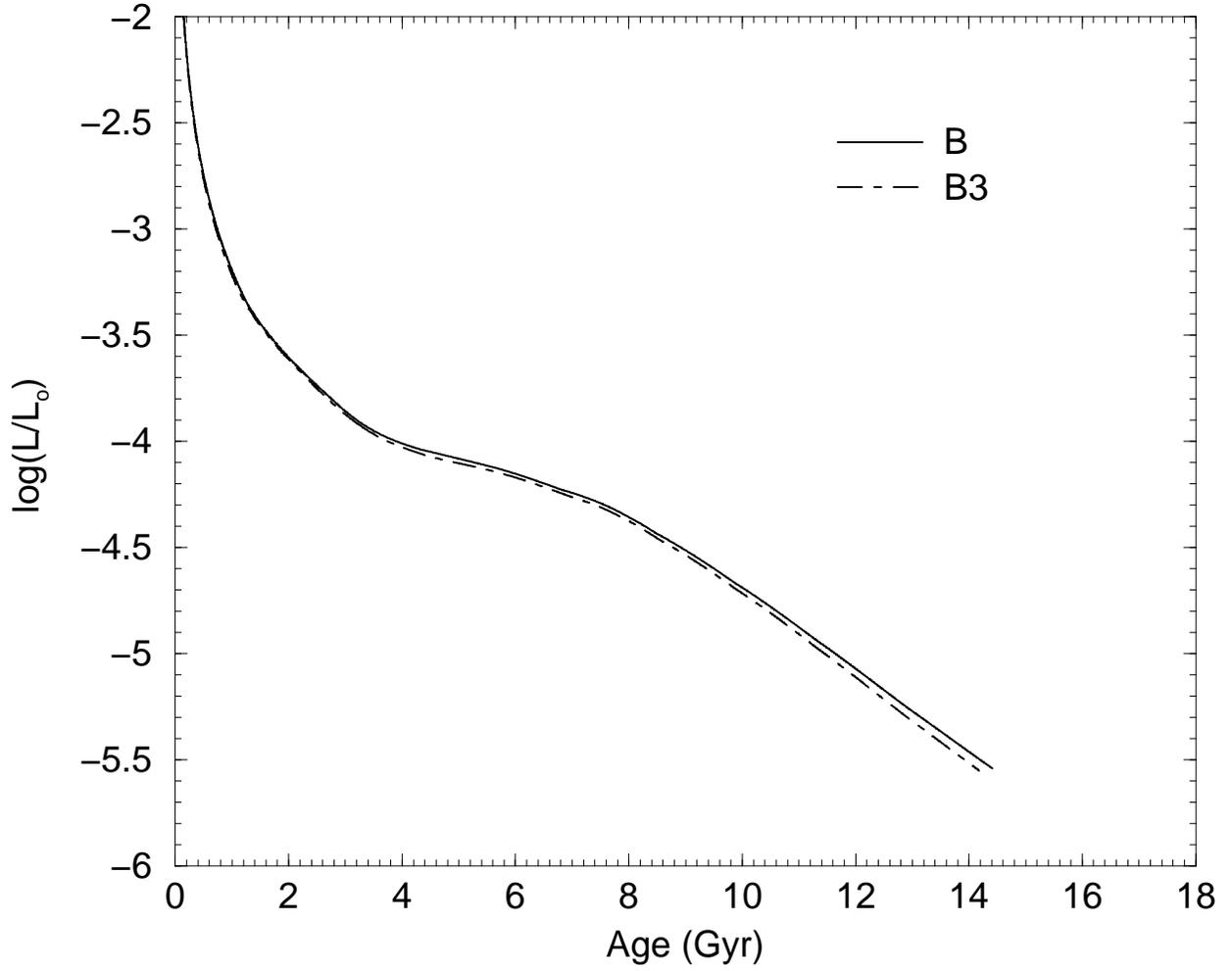}
\caption{Comparison between the cooling sequence obtained for a metal poor progenitor 
(model B with Z=$10^{-4}$) and a metal rich progenitor (model B3 with Z=0.02).\label{fig5}}
\end{figure}

\clearpage

\begin{figure}
\plotone{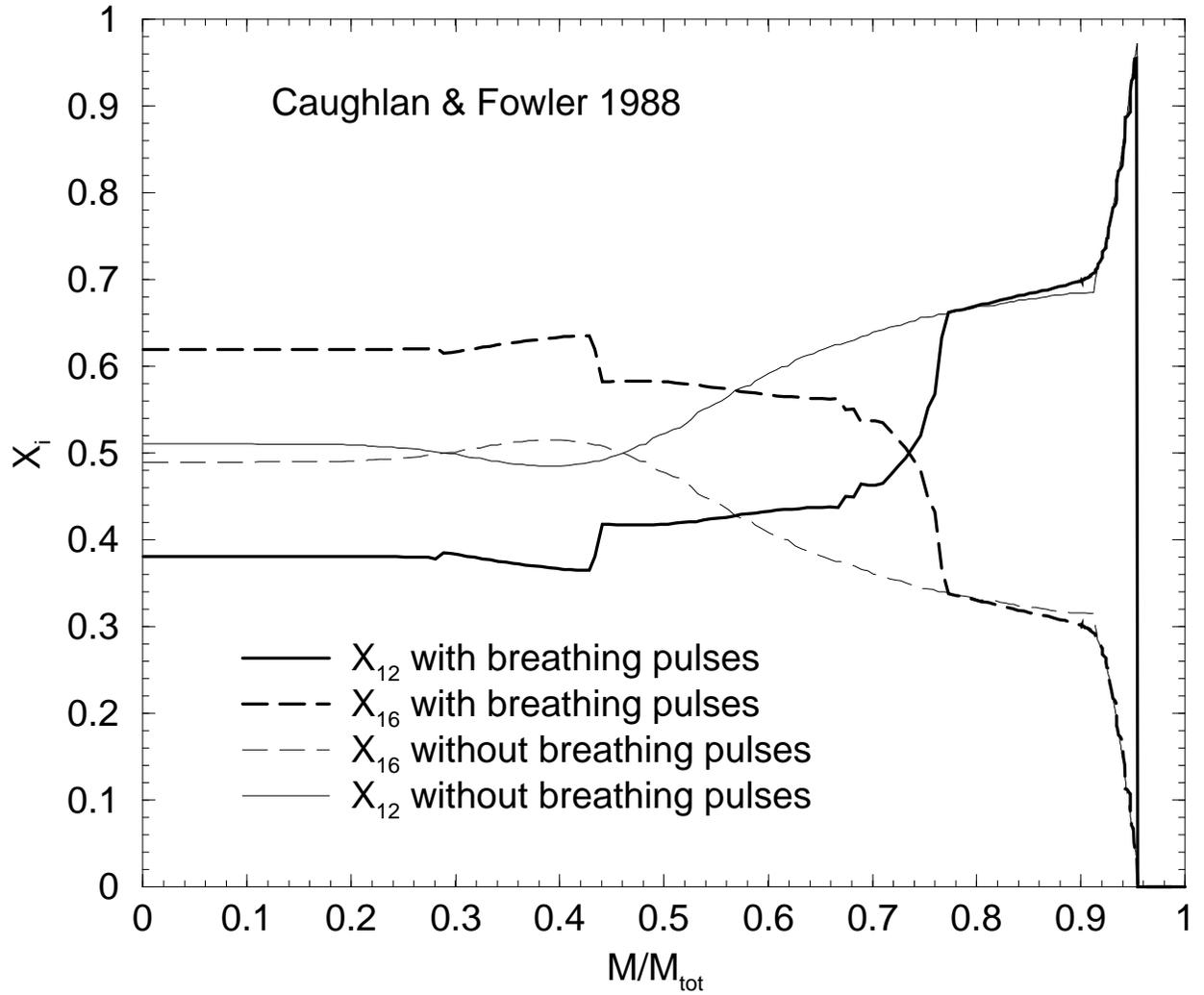}
\caption{Profiles of carbon (solid lines) and oxygen (dashed lines) for the model with
 (thick lines) and without (thin lines)  breathing pulses. In both cases, a 
low $^{12}$C$(\alpha,\gamma)^{16}$O rate (CF88) has been adopted.\label{fig6}}
\end{figure}

\clearpage

\begin{figure}
\plotone{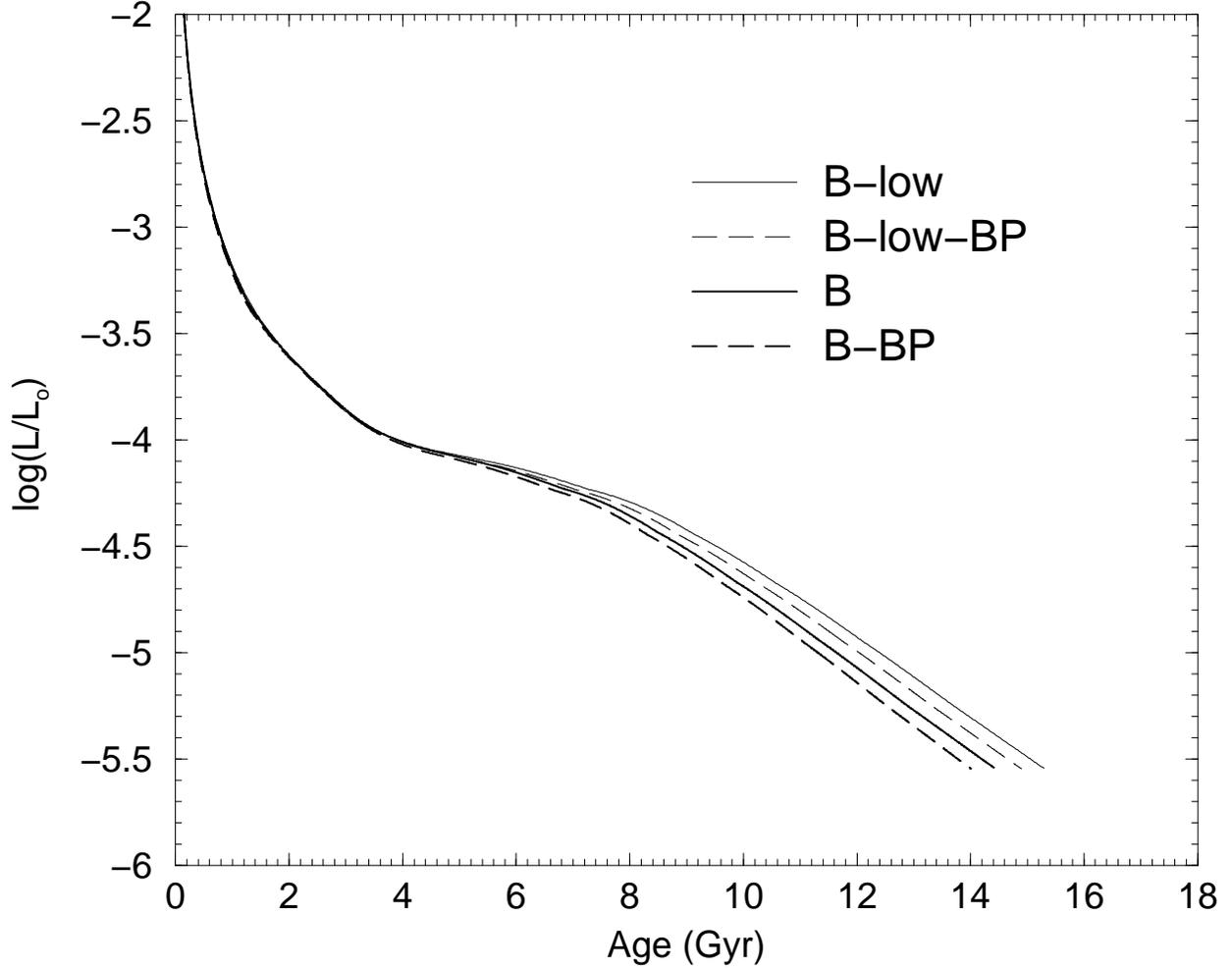}
\caption{Cooling sequences obtained from progenitor evolutions with and without breathing
pulses and different $^{12}$C$(\alpha,\gamma)^{16}$O rate: no-BPs and CF88 (thin solid line), 
BPs and CF88 (thin dashed line), no-BPs and CF85 thick solid line, BPs and CF85 
(thick dashed line). Note that
CF88 and CF85 are representative of a low and a high rate for the $^{12}$C$(\alpha,\gamma)^{16}$O
(see text).\label{fig7}}
\end{figure}

\clearpage

\begin{figure}
\plotone{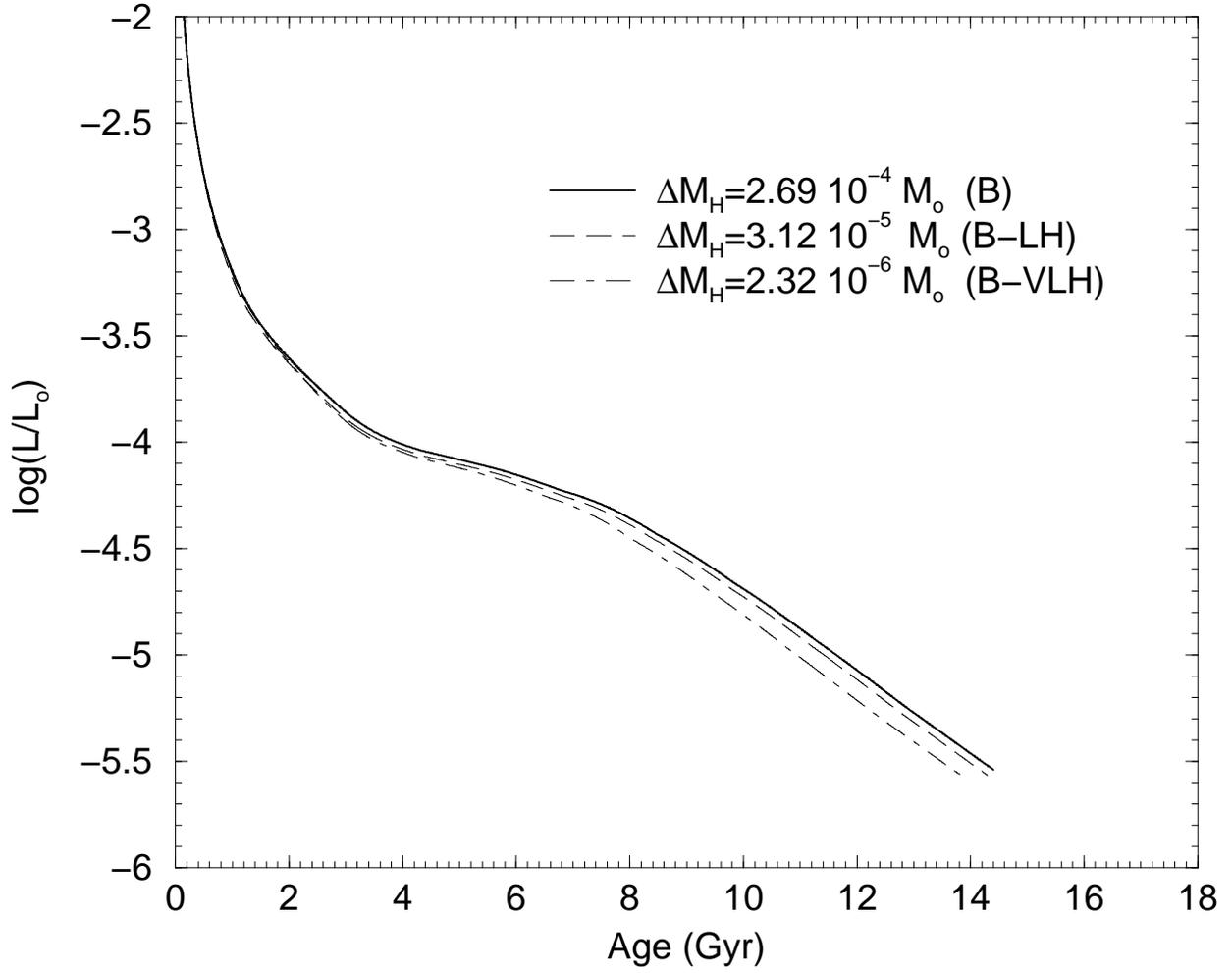}
\caption{Effects, on the cooling sequence, of the  
extension in mass of the H-rich envelope ($\Delta M_H$).\label{fig8}}
\end{figure}

\clearpage

\begin{figure}
\plotone{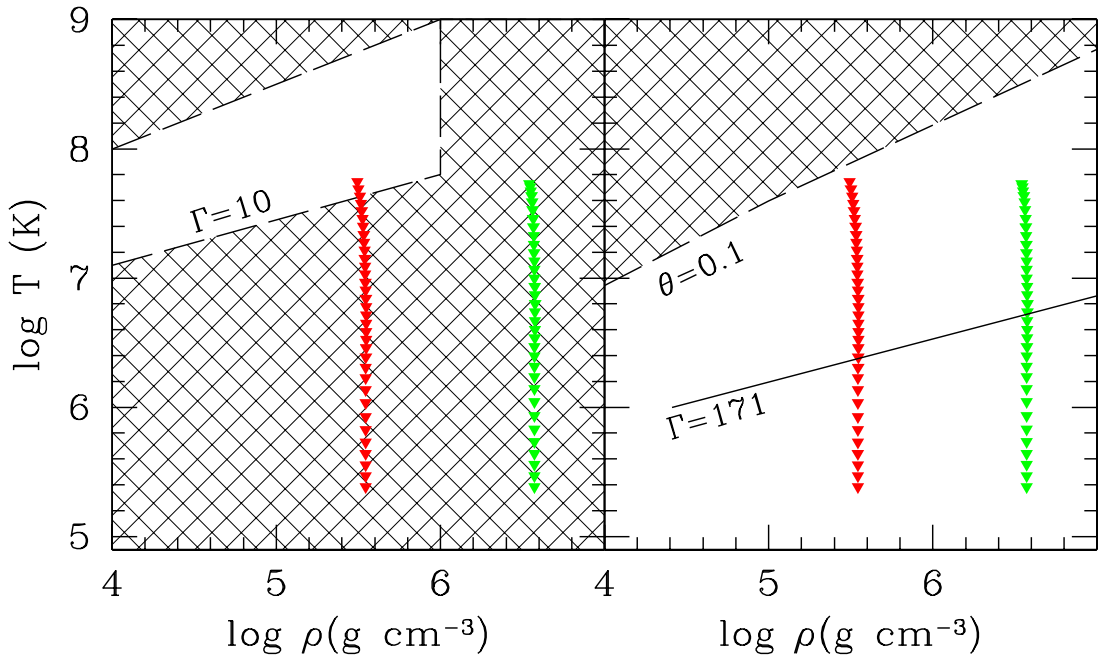}
\caption{Temperature-density diagram for the core. Left panel: hatched area shows where the HL69
conductive opacity are not valid. Right panel: the same but for the I93 conductive opacity.
The two series of arrows indicate the evolution of the central conditions and that of the outer border 
of the core. 
\label{fig9}}
\end{figure}

\clearpage

\begin{figure}
\plotone{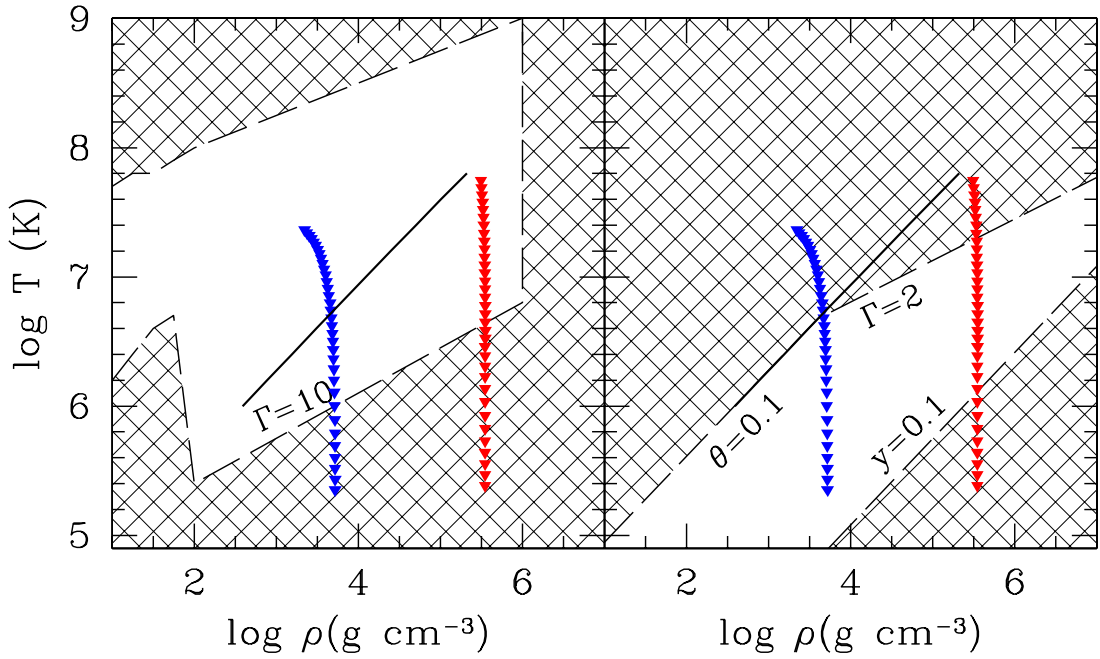}
\caption{Same as figure 9, but for the He-rich mantel. Here, the
two series of arrows indicate the evolutions of the physical conditions at the two  
boundary of the mantel. The thick solid line shows the location of the $\theta=0.1$
condition; model B-comb has been obtained by using HL69 above this line and I93 below.  
\label{fig10}}
\end{figure}

\clearpage

\begin{figure}
\plotone{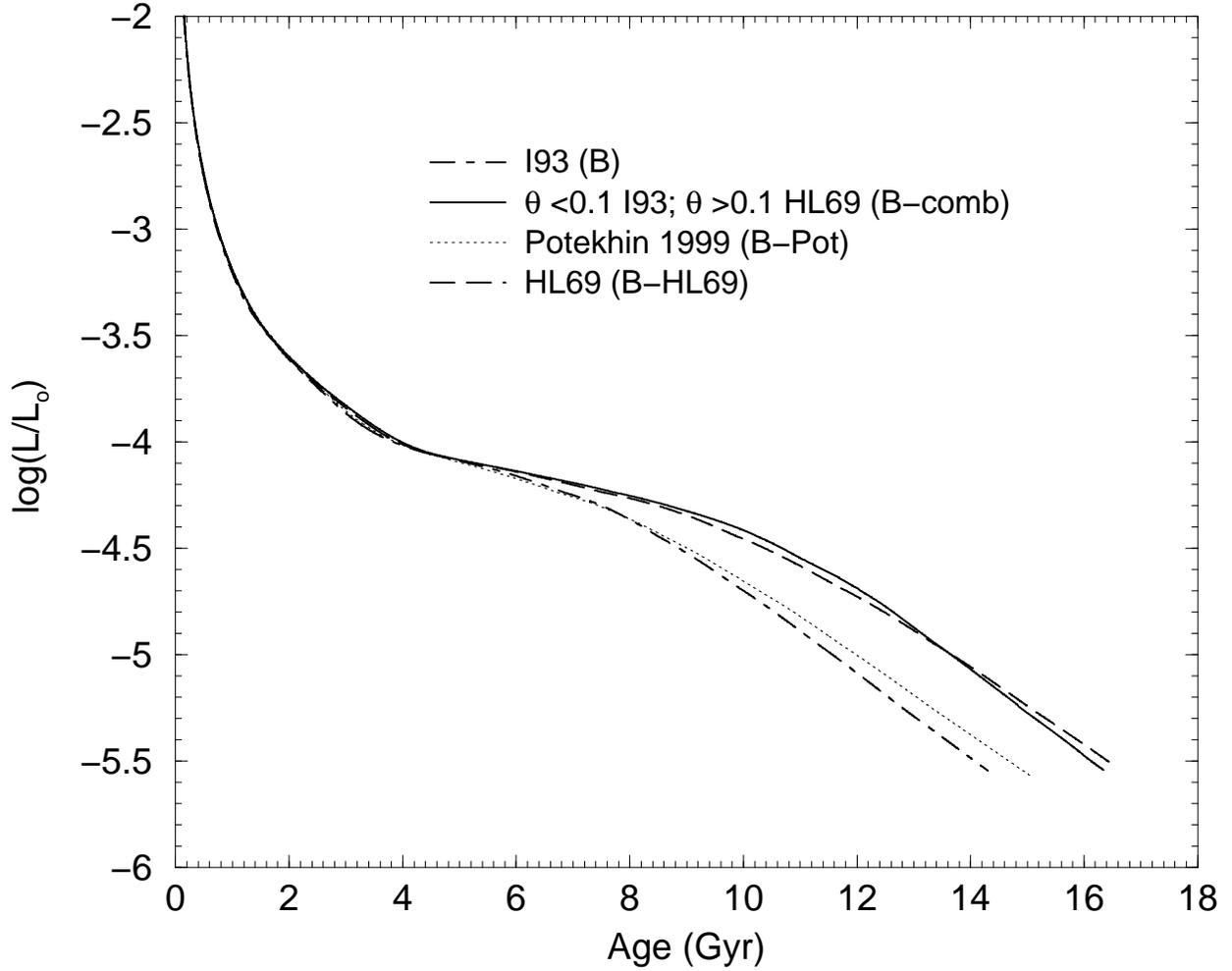}
\caption{Theoretical cooling sequences under different
prescriptions for the conductive opacity: I93, Mitake et al. 1984 
for the fluid phase and Itoh et al. 1993 for the solid phase (dot-dashed line); 
HL69, Hubbard \& Lampe 1969 (dashed line); I93 for the fully degenerate regime ($\theta \le 0.1$) and 
HL69 for the partially degenerate regime (solid line); 
Potekhin et al. 1999 (dotted line).\label{fig11}}
\end{figure}

\clearpage

\begin{figure}
\plotone{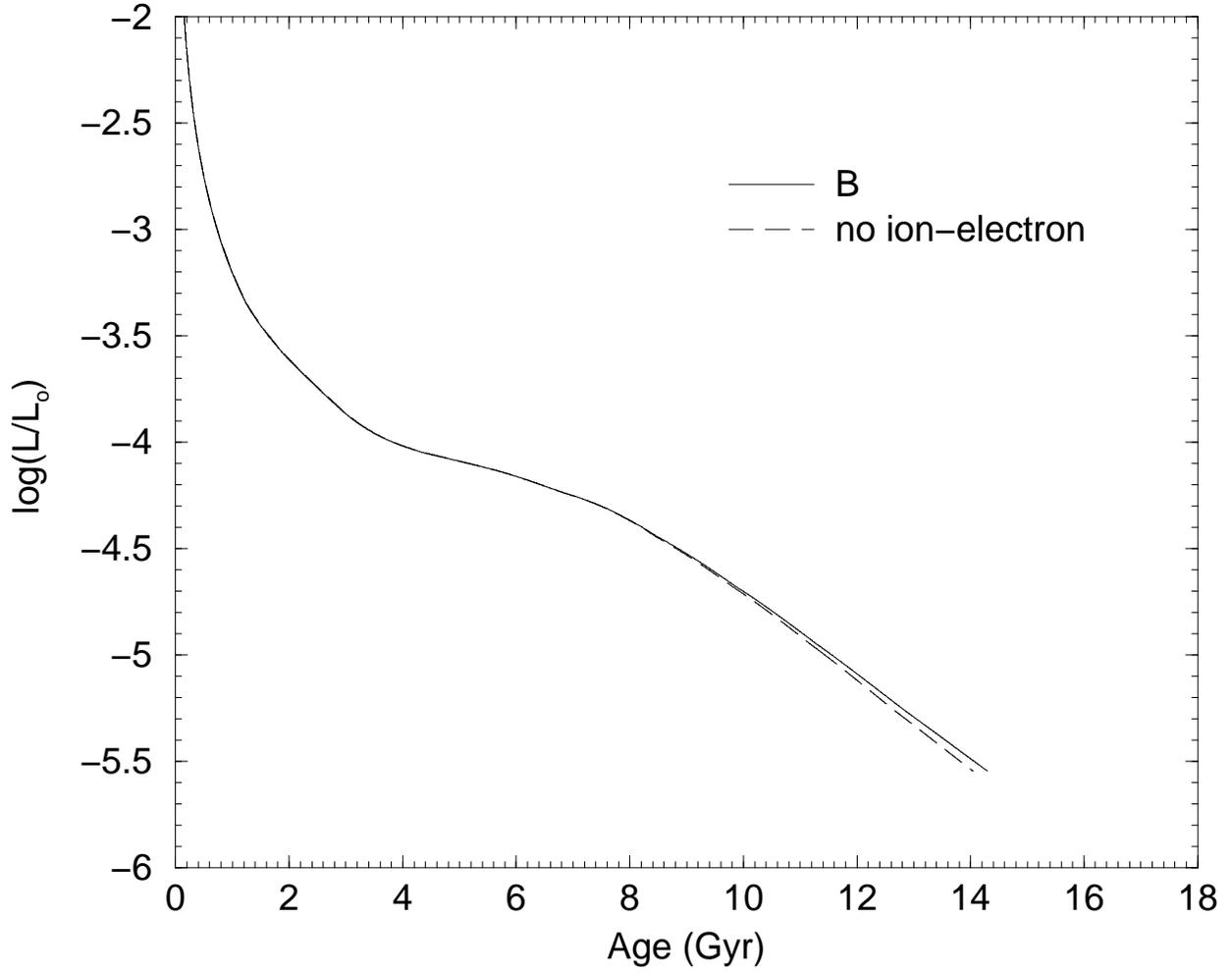}
\caption{Comparison between the cooling sequences computed adopting an EOS with 
(case B, solid line) 
and without (dashed line) the ion-electron interaction.\label{fig12}}
\end{figure}

\clearpage

\begin{figure}
\plotone{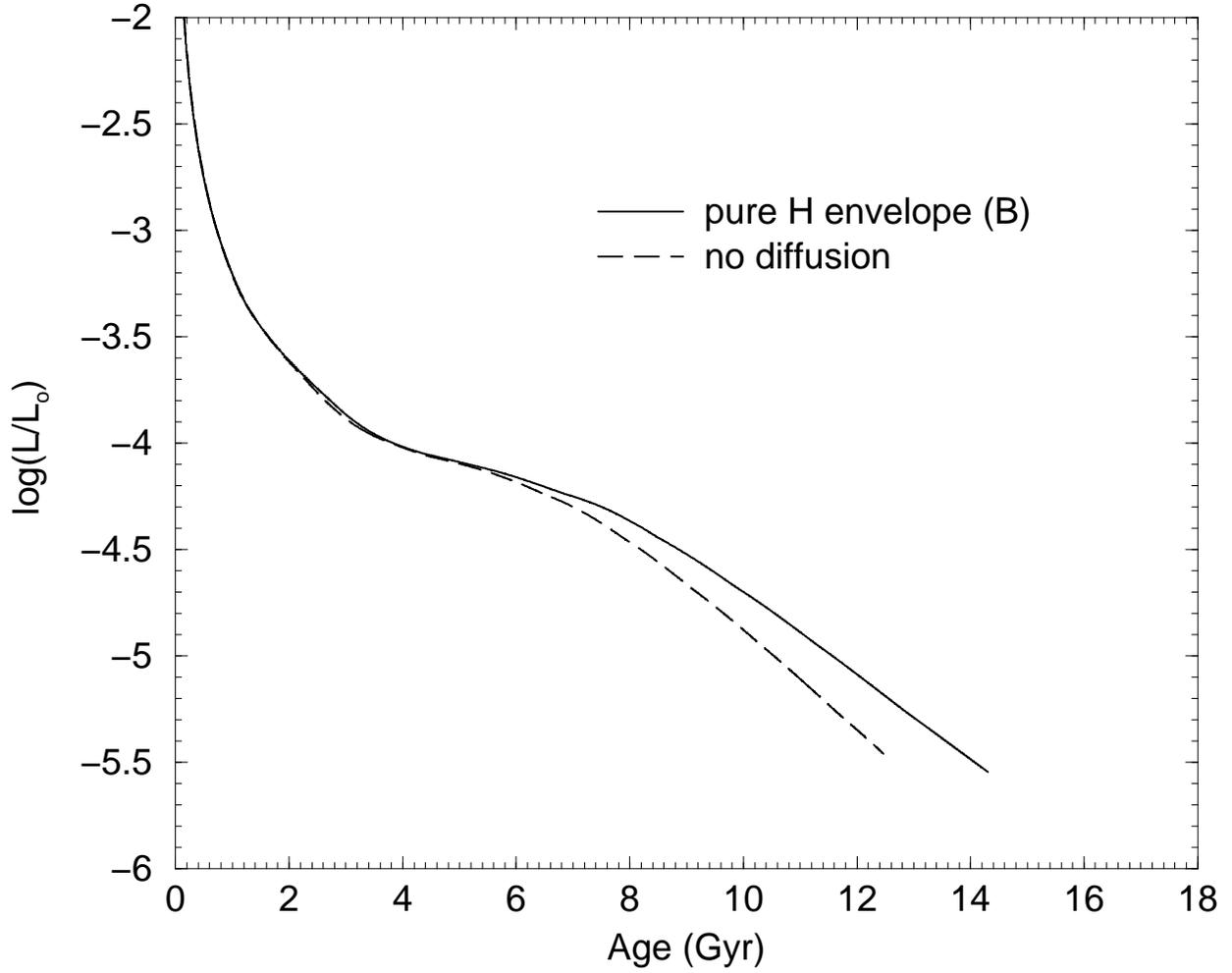}
\caption{Cooling sequences with (solid line) and without (dashed line) 
the chemical sorting of the H and He layers.\label{fig13}}
\end{figure}

\clearpage

\begin{figure}
\plotone{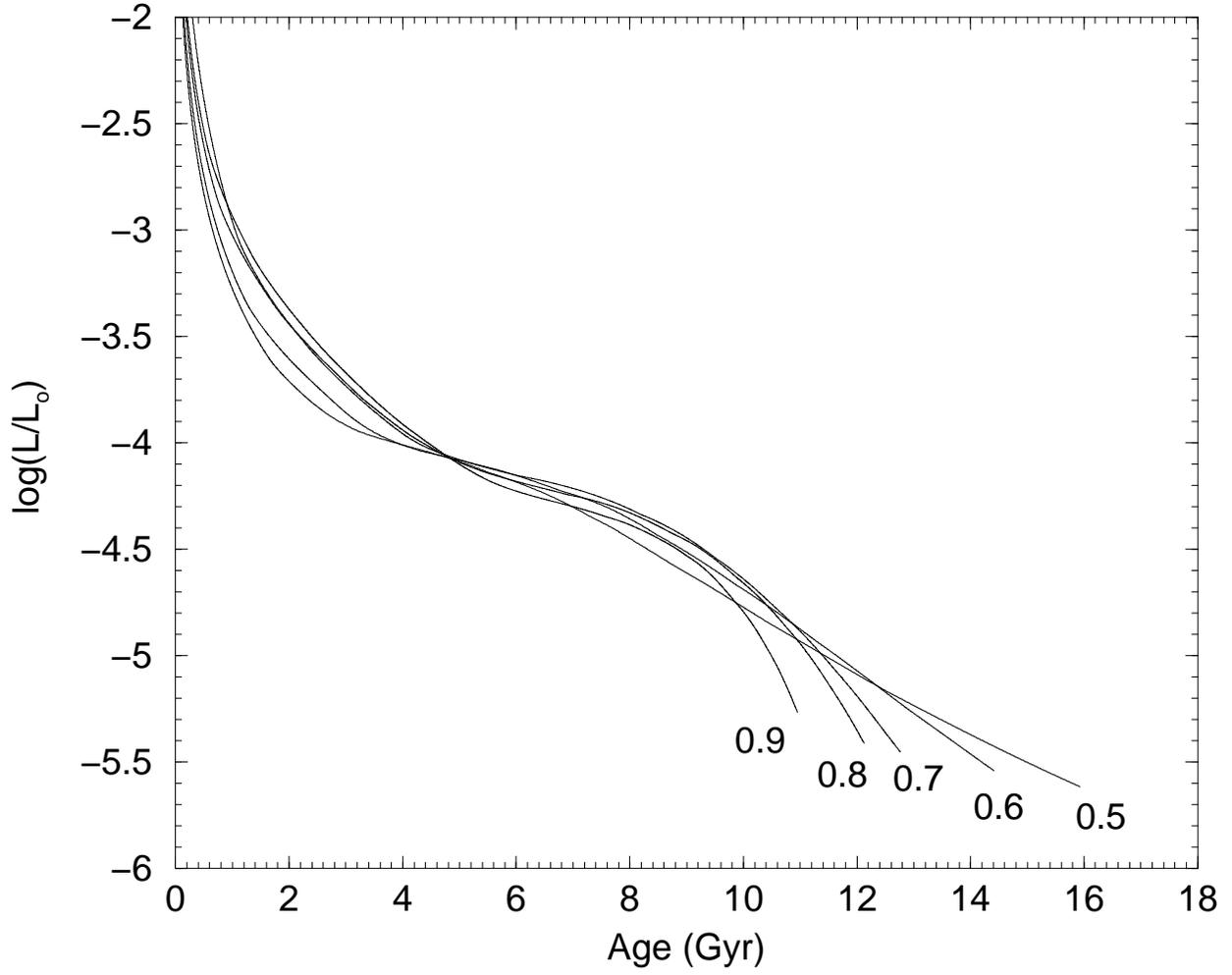}
\caption{Cooling sequences of WDs having mass ranging between 0.5 and 0.9 M$_\odot$.\label{fig14}}
\end{figure}


\clearpage

\begin{deluxetable}{cccccccccc} 
\tabletypesize{\scriptsize}
\tablecaption{0.6 $M_{\odot}$ WD: progenitors and cooling \label{tab1}}
\tablewidth{0pt}
\tablehead {\colhead{label} & \colhead{M$_{ZAHB}$\tablenotemark{1}} &
\colhead {Z\tablenotemark{1}} &
\colhead {M$_{He}$\tablenotemark{1}} & 
\colhead {M$_{WD}$\tablenotemark{2}} &
\colhead {$\Delta$ M$_{He}$\tablenotemark{2}} &
\colhead {$\Delta$ M$_{H}$ \tablenotemark{2}} &
\colhead {$X_{12}^c$\tablenotemark{3}} &
\colhead {Q\tablenotemark{4}} &
\colhead {$\Delta$t \tablenotemark{5}}}
\startdata
B & 0.600      &  0.0001  & 0.599  & 0.600 & $2.56 \times 10^{-2}$  & $2.69 \times 10^{-4}$ & 0.263 & 0.573 & $14.20 $ \\
B1 & 0.600     &  0.0001  & 0.580  & 0.600 & $1.29 \times 10^{-2}$  & $2.53 \times 10^{-4}$ & 0.258 & 0.542 & $13.74 $ \\  
B2 & 0.640     &  0.0001  & 0.515  & 0.601 & $2.03 \times 10^{-2}$  & $2.23 \times 10^{-4}$ & 0.249 & 0.516 & $14.05 $ \\
B3 & 0.600     &  0.0200  & 0.599  & 0.600 & $2.49 \times 10^{-2}$  & $8.48 \times 10^{-5}$ & 0.248 & 0.509 & $13.92 $ \\
B-LH & 0.600   &  0.0001  & 0.599  & 0.600 & $2.56 \times 10^{-2}$  & $3.12 \times 10^{-5}$ & 0.263 & 0.573 & $13.96 $ \\
B-VLH & 0.600  &  0.0001  & 0.599  & 0.600 & $2.56 \times 10^{-2}$  & $2.32 \times 10^{-6}$ & 0.263 & 0.573 & $13.49 $ \\
B-BP & 0.600   &  0.0001  & 0.599  & 0.600 & $2.54 \times 10^{-2}$  & $2.64 \times 10^{-4}$ & 0.151 & 0.316 & $13.78 $ \\
B-low & 0.600  &  0.0001  & 0.599  & 0.600 & $2.73 \times 10^{-2}$  & $2.82 \times 10^{-4}$ & 0.511 & 0.483 & $15.07 $ \\
B-low-BP & 0.600 & 0.0001 & 0.599  & 0.600 & $2.71 \times 10^{-2}$  & $2.72 \times 10^{-4}$ & 0.381 & 0.434 & $14.65 $ \\
B-HL69 & 0.600 &  0.0001  & 0.599  & 0.600 & $2.56 \times 10^{-2}$  & $2.69 \times 10^{-4}$ & 0.263 & 0.573 & $16.67 $ \\
B-comb & 0.600 &  0.0001  & 0.599  & 0.600 & $2.56 \times 10^{-2}$  & $2.69 \times 10^{-4}$ & 0.263 & 0.573 & $16.12 $ \\
B-Pot  & 0.600 &  0.0001  & 0.599  & 0.600 & $2.56 \times 10^{-2}$  & $2.69 \times 10^{-4}$ & 0.263 & 0.573 & $14.68 $ \\
B-noie & 0.600 &  0.0001  & 0.599  & 0.600 & $2.56 \times 10^{-2}$  & $2.69 \times 10^{-4}$ & 0.263 & 0.573 & $13.82 $ \\
B-nodif & 0.600 &  0.0001 & 0.599  & 0.600 & $2.56 \times 10^{-2}$  & $2.69 \times 10^{-4}$ & 0.263 & 0.573 & $12.63 $ \\ 
\enddata

\tablenotetext{1}{Start parameters (ZAHB): mass ($M_{ZAHB}[M_{\odot}]$), metallicity (Z), He core mass ($M_{He}[M_{\odot}]$)}
\tablenotetext{2}{WD parameters: total mass ($M_{WD}[M_{\odot}]$), He mass ($\Delta M_{He}[M_{\odot}]$) and H mass ($\Delta M_{H}[M_{\odot}]$)}
\tablenotetext{3}{Mass fraction of carbon at the center of the WD}
\tablenotetext{4}{Ratio between the mass of the innermost homogeneous region and the total WD mass}
\tablenotetext{5}{age (in Gyr) at the faint end, namely the time necessary to cool down from $log(L/L_{\odot})= 0$
 to $log(L/L_{\odot})=-5.5$.}
\end{deluxetable}

\clearpage

\begin{deluxetable}{ccccccc} 
\tabletypesize{\scriptsize}
\tablecaption{WD models with different mass \label{tab2}}
\tablewidth{0pt}
\tablehead {\colhead{label} &
\colhead {M$_{WD}$\tablenotemark{1}} &
\colhead {$\Delta$ M$_{He}$\tablenotemark{2}} &
\colhead {$\Delta$ M$_{H}$ \tablenotemark{3}} &
\colhead {$X_{12}^c$\tablenotemark{4}} &
\colhead {Q\tablenotemark{5}} &
\colhead {$\Delta$t \tablenotemark{6}}}
\startdata
A &  0.500 & $4.62 \times 10^{-2}$  & $4.24 \times 10^{-4}$ & 0.221  & 0.571 & 15.00  \\
B &  0.600 & $2.56 \times 10^{-2}$  & $2.69 \times 10^{-4}$ & 0.263  & 0.573 & 14.20  \\
C &  0.700 & $8.92 \times 10^{-3}$  & $1.27 \times 10^{-4}$ & 0.294  & 0.560 & 12.90  \\
D &  0.800 & $5.85 \times 10^{-3}$  & $6.15 \times 10^{-5}$ & 0.302  & 0.529 & 12.31  \\
E &  0.900 & $3.04 \times 10^{-3}$  & $3.87 \times 10^{-5}$ & 0.309  & 0.492 & 11.32  \\
\enddata

\tablenotetext{1}{WD mass ($M_{\odot}$)}
\tablenotetext{2}{He mass ($M_{\odot}$)}
\tablenotetext{3}{H mass  ($M_{\odot}$)}
\tablenotetext{4}{Mass fraction of carbon at the center of the WD}
\tablenotetext{5}{Ratio between the mass of the innermost homogeneous region and the total WD mass}
\tablenotetext{6}{age (in Gyr) at the faint end, namely the time necessary to cool down from $log(L/L_{\odot})= 0$
 to $log(L/L_{\odot})=-5.5$.}
\end{deluxetable}


\begin{thebibliography}{}
\bibitem[]{1} Alexander D.R., \& Ferguson J.W. 1994, \apj,  437, 879
\bibitem[]{2} Althaus, L.G., Serenelli, A.M., Corsico, A.H., \& Benvenuto, O.G. 2002, \mnras,  330, 685
\bibitem[]{3} Angulo, C. et al. 1999, Nucl. Phys. A, 656, 3
\bibitem[]{4} Benvenuto, O.G., \& Althaus, L.G. 1999, \mnras, 303, 30
\bibitem[]{5} Bergeron, P.  2000, private comunication
\bibitem[]{6} Bergeron, P., Ruiz, M.T., \& Leggett, S.K. 1997, \apjs, 108, 339
\bibitem[]{7} Bergeron, P., Saumon, D., \& Wesemael, F. 1995, \apj,  443, 764
\bibitem[]{8} Bergeron, P., Wesemael, F., \& Beauchamp, A. 1995, PASP,  107, 1047
\bibitem[]{9} Bergeron, P., Wesemael, F., \& Fontaine, G. 1991, \apj,  367, 253
\bibitem[]{10} Bradley, P.A., \& Winget, D.E. 1991, \apjs,  75, 463
\bibitem[]{11} Bradley, P.A., Winget, D.E., \& Wood, M.A. 1992, \apj,  391, L33
\bibitem[]{12} Buchmann, L. 1996, \apj, 468L, 127
\bibitem[]{13} Buchmann, L. 1997, \apj, 479L, 153
\bibitem[]{14} Cassisi, S., Castellani, V., \& Tornamb\'e, A. 1996, \apj,  459, 298
\bibitem[]{15} Castellani, M., Limongi, M., \& Tornamb\'e, A. 1995, \apj, 450, 275
\bibitem[]{16} Castellani, V., Chieffi, A., Tornamb\'e, A., \& Pulone, L. 1985, \apj,  296, 204
\bibitem[]{17} Castellani, V., Degl'Innocenti, S., \& Romaniello, M. 1994, \apj, 423, 266
\bibitem[]{18} Caughlan, G.R., \& Fowler, W.A. 1988, Atomic Data Nucl. Data Tables, 40, 283
\bibitem[]{19} Caughlan, G.R.,  Fowler, W.A., Harris, M.J., \& Zimmermann, B.A. 1985, Atomic Data Nucl. Data Tables, 32, 197
\bibitem[]{20} Chabrier, G., Ashcroft, N.W., \& De Witt, H.E. 1992, Nat., 360, 48
\bibitem[]{21} Chabrier, G., Brassard, P., Fontaine, G., \& Saumon, D. 2000, \apj, 543, 216
\bibitem[]{22} Chieffi A., \& Straniero O. 1989, \apj,  71, 47
\bibitem[]{23} Clemens, J.C. 1993, Baltic. Astron., 2, 407
\bibitem[]{24} Cohen, E.G.D., \& Murphy, T.J. 1969, Phys. Fluids, 12, 1404
\bibitem[]{25} D'Antona, F., \& Mazzitelli, I. 1989, \apj, 347, 934
\bibitem[]{26} D'Antona, F., \& Mazzitelli, I. 1990, ARA\&A, 28, 139
\bibitem[]{27} De Witt, H.E., Graboske, H.C., \& Cooper, M.S. 1973, \apj,  181, 439
\bibitem[]{28} De Witt, H., \& Slattery, W. 1999, Contrib. Plasma Phys., 39, 97
\bibitem[]{29} Dominguez, I., Chieffi, A., Limongi, M., \& Straniero, O. 1999, \apj,  524, 226
\bibitem[]{30} Dominguez, I., H\"oflich, P., \& Straniero, O. 2001, \apj,  557, 279
\bibitem[]{31} Dominguez, I., Straniero, O., Tornamb\'e, A., \& Isern, J. 1996, \apj, 472, 783
\bibitem[]{32} Dominguez, I., Tornamb\'e, A., \& Isern, J. 1993, \apj, 419, 268
\bibitem[]{33} Farouki, R.T., \& Hamaguchi, S. 1993, Phys. Rev. E., 47, 4330
\bibitem[]{34} Fontaine, G., \& Michaud, G. 1979, \apj, 231, 826
\bibitem[]{35} Fujimoto, M.Y. 1982a, \apj, 257, 767 
\bibitem[]{36} Fujimoto, M.Y. 1982b, \apj, 257, 752
\bibitem[]{38} Garcia-Berro, E., Ritossa, C., \& Iben, I.Jr. 1997, \apj, 485, 765
\bibitem[]{39} Gialanella et al. 2001, Eur. Phys. J.A., 11, 357
\bibitem[]{40} Graboske, H., De Witt, H., Grossman, A., \& Cooper M. 1973, \apj, 181, 457
\bibitem[]{41} Haft, M., Raffelt, G., \& Weiss A. 1994, \apj, 425, 222
\bibitem[]{42} Habing, H.J. 1996, A\&A Rev, 7, 97
\bibitem[]{43} Hansen, B.M. 1998, Nat., 394, 860
\bibitem[]{44} Hansen, B.M. 1999, \apj, 520, 680
\bibitem[]{45} Herwig, F. 1997, in Proceedings of the 10th European Workshop on White Dwarfs, held in Blanes,
                Spain, 17-21 June 1996. Ed. J. Isern, M. Hernanz, and E. Gracia-Berro. Publisher: Dordrecht:
                Kluwer Academic Publishers, 1997
\bibitem[]{46} Hubbard, W.B., \& Lampe, M. 1969, \apjs, 18, 297
\bibitem[]{47} Iben, I.J., 1972, \apj, 178, 433 
\bibitem[]{48} Iben, Jr.I. 1982, \apj, 259, 244
\bibitem[]{49} Iben, Jr.I., \& MacDonald, J. 1985, \apj, 296, 540
\bibitem[]{50} Iben, Jr.I., \& MacDonald, J. 1986, \apj, 301, 164
\bibitem[]{51} Iben, Jr.I., \& Renzini, A. 1983, ARA\&A, 21, 271
\bibitem[]{52} Iben, Jr.I., \& Tutukov, A.V. 1986, \apj, 311, 742
\bibitem[]{53} Iglesias, A., \&  Rogers, F.J. 1996, \apj, 464, 943 
\bibitem[]{54} Imbriani, G., Limongi, M., Gialanella, L., Terrasi, F., Straniero, O., \& Chieffi, A. 2001, \apj, 558, 903
\bibitem[]{55} Isern, J., Garcia-Berro, E., Hernanz, M., \& Chabrier, G. 2000, \apj, 528, 397
\bibitem[]{56} Isern, J., Mochkovitch, R., Garcia-Berro, E., \& Hernanz, M. 1997, \apj, 485, 308
\bibitem[]{57} Itoh, N., Adachi, T., Nakagawa, M., Kohyama, Y., \& Munakata, H. 1989, \apj, 339, 354
\bibitem[]{58} Itoh N., Hayashi, H., \& Kohyama, Y. 1993, \apj, 418, 405
\bibitem[]{59} Itoh N., Mitake S., Iyetomi H., \& Ichimaru S., 1983, \apj, 273, 774
\bibitem[]{60} Itoh, N., Totsuji, H., \& Ichimaru, S. 1977, \apj, 218, 477
\bibitem[]{61} Itoh, N., Totsuji, H., Ichimaru, S., \& De Witt, H. 1979, \apj, 234, 1079
\bibitem[]{62} Kippenhahn, R., Kohl, K., \& Weigert, A. 1967, Zeitschrift fur Astrophysik, 66, 58
\bibitem[]{63} Kippenhahn, R., Thomas, H.C., \& Weigert, A. 1968, Zeitschrift fur Astrophysik, 69, 265
\bibitem[]{64} Kunz R., Jaeger F.M., Mayer A., \& Hammer J.W. 2001, Phys. Rev. Lett., 86, 3244
\bibitem[]{65} Kunz R., Jaeger F.M., Mayer A., Hammer J.W. , Staudt, G., Harissopulos, S., \& Paradellis, T. 2002, \apj, 567, 643
\bibitem[]{66} Lamb, D. Q., \& Van Horn, H. M. 1975, \apj, 200, 306
\bibitem[]{67} Landau, L.D., \& Lifshitz, E.M. 1969, "Statistical Physics", Pergamon, Oxford 
\bibitem[]{68} Livne, E., \& Tuchman, Y. 1988, \apj, 332, 271
\bibitem[]{69} Mazzitelli, I. 1994, ``The equation of state in astrophysics'', Proceedings of IAU Colloquium No. 147, Saint-Malo,
                France, 14-18 June 1993, Cambridge: Cambridge University Press, 1994, ed. G. Chabrier and E. Schatzman, p.144
\bibitem[]{669} Mestel, L. 1952, \mnras, 112, 583
\bibitem[]{70} Metcalfe, T.S., Winget, D.E., \& Charbonneau, P. 2001, \apj, 557, 1021
\bibitem[]{71} Mitake, S., Ichimaru, S., \& Itoh, N. 1984, \apj, 277, 375
\bibitem[]{72} Montgomery, M.H., Klumpe, E.W., Winget, D. E., \& Wood, M.A. 1999, \apj, 525, 482
\bibitem[]{73} Muchmore, D. 1984, \apj, 278, 769
\bibitem[]{74} Paquette, C., Pelletier, C., Fontaine, G., \& Michaud, G. 1986, \apjs, 61, 197
\bibitem[]{75} Paresce, F., De Marchi, G., \& Romaniello, M. 1995, \apj, 440, 216
\bibitem[]{76} Piersanti, L., Cassisi, S., Iben, I.Jr., \& Tornamb\'e, A. 2000, \apj, 535, 932
\bibitem[]{77} Potekhin, A.Y. 1999, A\&A, 351, 787
\bibitem[]{78} Potekhin, A.Y., Baiko, D.A., Haensel, P., \& Yakovlev, D.G. 1999, A\&A, 346, 345
\bibitem[]{79} Potekhin, A.Y., \& Chabrier, G. 2000, Phys. Rev. E., 62, 8554
\bibitem[]{80} Richer, H.B. et al. 1997, \apj, 484, 741
\bibitem[]{81} Richer, H.B. et al. 2000, \apj, 529, 318 
\bibitem[]{82} Ritossa, C., Garcia-Berro, E., \& Iben, Jr.I 1996, \apj, 460, 489
\bibitem[]{83} Ritossa, C., Garcia-Berro, E., \& Iben, Jr.I 1999, \apj, 515, 381
\bibitem[]{84} Salaris, M., Dominguez, I., Garcia-Berro, E., Hernanz, M., Isern, J., \& Mochkovitch, R. 1997, \apj, 486, 413
\bibitem[]{85} Salaris, M., Garcia-Berro, E., Hernanz, M., Isern, J., \& Saumon, D. 2000, \apj, 544, 1036
\bibitem[]{86} Salpeter, E.E. 1961, \apj, 134, 669
\bibitem[]{87} Saumon, D., Chabrier, G., \& Van Horn, H.M. 1995, \apjs, 99, 713
\bibitem[]{88} Saumon, D., \& Jacobson, S.B. 1999, \apj, 511, L107
\bibitem[]{90} Segretain, L., Chabrier, G., Hernanz, M., Garcia-Berro, E., Isern, J., \& Mochkovitch, R. 1994, \apj, 434, 641
\bibitem[]{91} Stolzmann, W., \& Blocker, T. 2000, A\&A, 361, 115
\bibitem[]{92} Straniero, O. 1988, A\&A, 76, 157
\bibitem[]{93} Straniero, O., Chieffi, A., \& Limongi, M. 1997, \apj, 490, 425
\bibitem[]{94} Straniero, O., Chieffi, A., Limongi, M., Busso, M., Gallino, R., \& Arlandini, C. 1997, \apj, 478, 332
\bibitem[]{95} Tanaka, S., Mitake, S., \& Ichimaru, S. 1985, Phys. Rev. A, 32, 1896
\bibitem[]{96} Tassoul, M., Fontaine, G., \& Winget, D.E. 1990, \apjs, 72, 335
\bibitem[]{97} Von Hippel, T., Gilmore, G., \& Jones, D.H.P. 1995, \mnras, 273L, 39 
\bibitem[]{98} Von Hippel, T., \& Gilmore, G. 2000, \aj, 120, 1384
\bibitem[]{99} Weidemann, V. 1987, A\&A, 188, 74
\bibitem[]{100} Weidemann, V. 2000, A\&A, 363, 647
\bibitem[]{101} Weidemann, V., \& Koester, D. 1983, A\&A, 121, 77
\bibitem[]{102} Wood, M.A. 1995, in ``White Dwarfs'', Springer Verlag, ed: D. Koester \& K. Werner, 41
\bibitem[]{103} Yakovlev, D., \& Shalybkov, D. 1989, Astron. Space Phys. Rev., 7, 311

\end{thebibliography}
\end{document}